%% file: main.tex
\documentclass[amsmath,amssymb,aps,prl,twocolumn,superscriptaddress]{revtex4-2}
\usepackage{amsmath}
\usepackage{tikz}
\tikzset{>=latex}

\usepackage{mathtools}
\usepackage{physics}
\usepackage{graphicx}
\usepackage{subcaption}
\usepackage{comment}
\usepackage[percent]{overpic}
\usepackage{bm}
\usepackage{svg}
\usepackage[colorlinks,pdfusetitle,urlcolor=blue,citecolor=blue,linkcolor=blue,bookmarksnumbered,plainpages=false]{hyperref}

\usepackage{pgfplots}

\pgfplotsset{compat=1.10}

\usepgfplotslibrary{fillbetween}
\usetikzlibrary{patterns}
\tikzset{>=latex}
\newcommand{\tens}{\mathbb}
\renewcommand{\vec}{\mathbf}

\usepackage[normalem]{ulem}

\begin{document}

\title{Engineering dipole-dipole couplings for enhanced cooperative light-matter interactions}

\author{Adam Burgess}%
\affiliation{SUPA, Institute of Photonics and Quantum Sciences, Heriot-Watt University, Edinburgh, EH14 4AS, UK}
\author{Madeline C.~Waller}%
\affiliation{School of Physics and Astronomy, University of Glasgow, Glasgow, G12 8QQ UK}
\author{Erik M.~Gauger}%
\affiliation{SUPA, Institute of Photonics and Quantum Sciences, Heriot-Watt University, Edinburgh, EH14 4AS, UK}
\author{Robert Bennett}%
\affiliation{School of Physics and Astronomy, University of Glasgow, Glasgow, G12 8QQ UK}


\date{\today}
\begin{abstract}
Cooperative optical effects are enabled and controlled by interactions between molecular dipoles, meaning that their mutual orientation is of paramount importance to, for example, superabsorbing light-harvesting antennas. Here we show how to move beyond the possibilities of simple geometric tailoring, demonstrating how a metallic sphere placed within a ring of parallel dipoles engineers an effective Hamiltonian that generates “guide-sliding” states within the ring system. This allows steady-state superabsorption in noisy room temperature environments, outperforming previous designs while being significantly simpler to implement. 
As exemplified by this showcase, our approach represents a powerful design paradigm for tailoring cooperative light-matter effects in molecular structures that extends beyond superabsorbing systems, to a huge array of quantum energy transport systems.

\end{abstract}

\maketitle
 

Dipolar interactions underpin countless important physical \cite{stuhlerObservationDipoleDipoleInteraction2005,zhangVisualizingCoherentIntermolecular2016,xuDipoleDipoleInteraction2024}, chemical \cite{xuTwoDimensionalSelfAssembledMolecular2012,kamerIntimateInteractionsCarbonyl2013} and biological phenomena \cite{wohlertRangeShieldingDipoleDipole2004,pauliniOrthogonalMultipolarInteractions2005,Plenio2017,Plenio2020}, with their influence extending to a wide variety of photophysics \cite{hestandExpandedTheoryJMolecular2018,spanoSpectralSignaturesFrenkel2010}. Acting between molecules possessing dipole moments they originate from Coulomb interactions of the charged constituents of each (overall neutral) molecule, exhibiting rich orientation-dependence. This leads to notably different couplings in structures whose molecules are aligned `face-to-face' (known as H-aggregates), distinct from those in the `head-to-tail' configuration (J-aggregates) \cite{hestandExpandedTheoryJMolecular2018}.  

One prominent contemporary use of the dipole-dipole interaction is the enhancement of effective optical absorption rates, as motivated by Dicke's prediction of superradiant atomic dipoles~\cite{Dicke1954,Sitek2007, Kozub2012,Celardo2012,Luo2019,Reitz22}. The concept of the time-reversed process --- superabsorption --- achievable by manipulating symmetric dipolar interactions to sustain a superabsorbing state has  recently gained considerable interest in the field  theoretically ~\cite{Higgins2014,Brown2019} and circumventing the need for dipolar interactions also experimentally~\cite{Yang:2021aa,Quach2022}. The concept draws inspiration from photosynthetic systems where J-aggregate-like ring arrangement reduces light-matter coupling due to destructive interference of the transition dipoles in the plane of the ring and can prevent damage by limiting solar absorption~\cite{Hu1997} and facilitate efficient long-range exciton transport through dark states~\cite{Mattioni21}. Conversely, artificial systems arranged as H-aggregates promise enhanced light-harvesting performance for improved solar energy conversion~\cite{Creatore2013,Zhang2015, Zhang2016,Fruchtman2016, Scholes2017,Romero2017,Higgins:2017aa,Potocnik:2018aa,Hu2018,Rouse_2019,Werren2023}.

A question then naturally presents itself --- is it possible to have the `best of both worlds', where, for example, an H aggregate with desirable far-field properties is imbued with near-field couplings associated with a J aggregate? Here we answer this question positively, exploiting that textbook dipolar interactions are not immutable laws of nature but are instead a particular case of a broader class 
of interactions in dielectric or metallic environments \cite{urbakhDipoledipoleInteractionsInterfaces1993,jonesModifiedDipoledipoleInteraction2018}. This is the realm of macroscopic QED \cite{grunerGreenfunctionApproachRadiationfield1996,scheelMacroscopicQEDConcepts2009}, where techniques originating in the study of dispersion forces \cite{casimirAttractionTwoPerfectly1948,casimirInfluenceRetardationLondonvan1948,buhmann2012Book1,buhmann2012Book2} can be used to engineer interatomic interaction processes almost at will, most recently 
in the rapidly-expanding field of inverse design \cite{bennettInverseDesignLight2020}. We then exploit these engineered dipole-dipole couplings and show that they enable the design of a high-performance light-harvesting superabsorber, providing a unique synthesis of macroscopic QED and quantum heat-engine formalisms.

\begin{figure}
    \centering
    \begin{overpic}[width=0.9\linewidth]{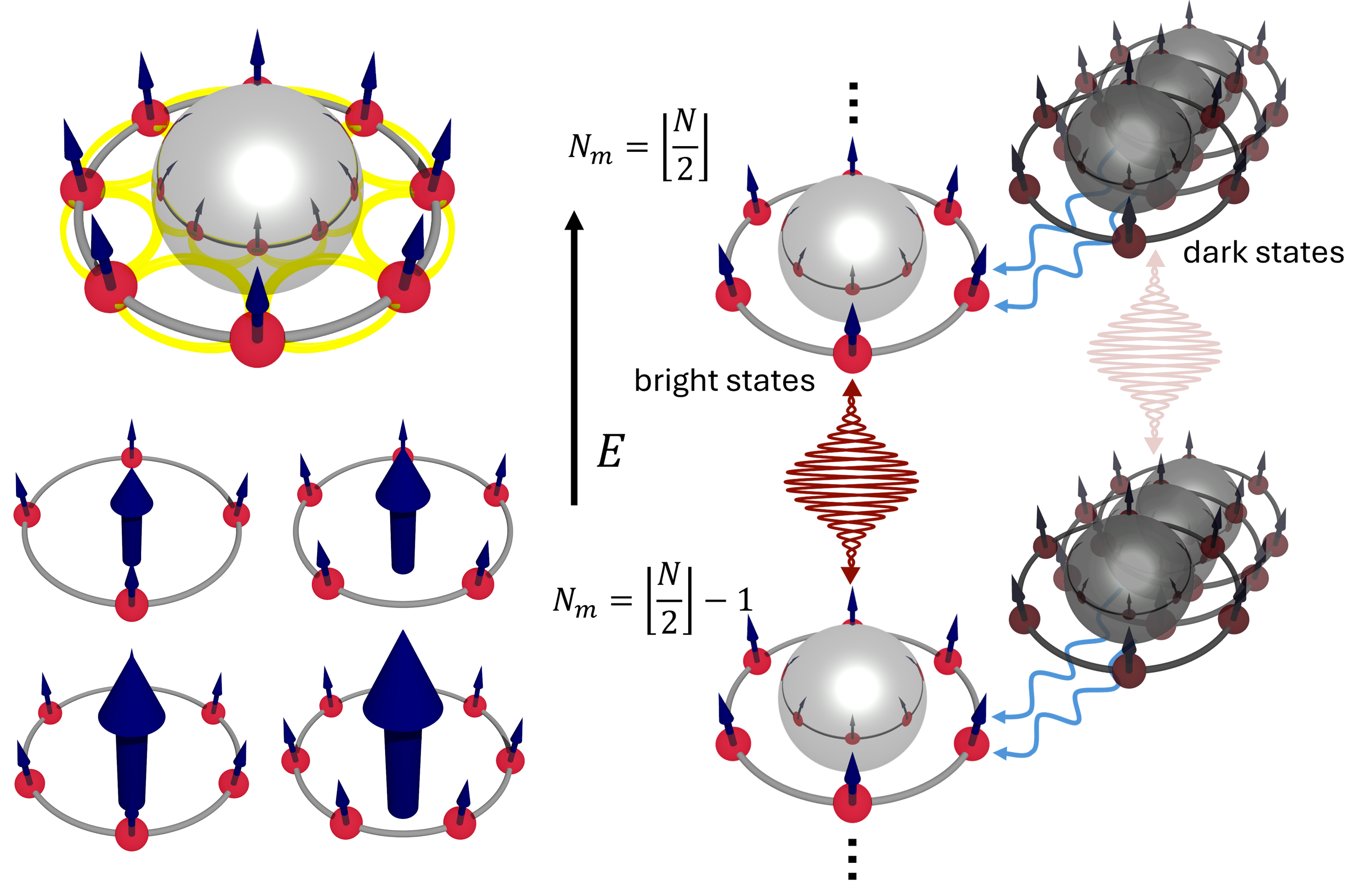}
 \put (2,65) {(a)}
 \put (45,65) {(b)}
 \put (2,35) {(c)}
\end{overpic}
     \caption{(a) A rendering of the parallel guide-slide super absorber, the placement of a dielectric sphere in the centre of the ring engineers the dipole-dipole couplings. (b) In turn these lead to a level structure that `guides' excitations onto optically brighter (approximately Dicke-ladder) states via phonon relaxation. (c) Parallel dipoles exhibit $N^2$ enhancement of the transition dipole strength near Dicke-ladder half-inversion $N_m = \lfloor\frac{N}{2}\rfloor$, allowing for superabsorption.}
    \label{fig:GuideSlideSchematic}
\end{figure}

The electrostatic limit of the free space coupling $J_{ij}^{(0)}$ between two dipoles $\vec{d}_i$ and $\vec{d}_j$ at positions $\vec{r}_i$ and $\vec{r}_j$ separated by $\bm{R}=\vec{r}_i-\vec{r}_j$ is
\begin{equation}
    J_{ij}^{(0)}= \frac{1}{4\pi |\bm{R}|^3}  \left[\vec{d}_i \cdot \vec{d}_j- 3 (\vec{d}_i \cdot \mathbf{e}_R) (\vec{d}_j \cdot \mathbf{e}_R \right)] , \label{J0MainText}
\end{equation}
%
%
where $\mathbf{e}_R = \bm{R}/|\bm{R}|$ 
and we have used a system of natural units with $c = \hbar = \varepsilon_0 =  1$. The generalisation of Eq.~\eqref{J0MainText} 
to an arbitrary environment and separations is \cite{dungResonantDipoledipoleInteraction2002,cortesSuperCoulombicAtomAtom2017}:
\begin{equation}\label{JijGeneral}
    J_{ij} = -\omega^2 \vec{d}_i \cdot \mathrm{Re}[ \tens{G}(\vec{r}_i,\vec{r}_j,\omega)] \cdot \vec{d}_j ,
\end{equation}
where $\omega$ is the average of the transition frequencies of the two dipoles (here assumed identical), and $\tens{G}$ is the dyadic Green's tensor that solves the Helmholtz equation with boundary conditions determined by the geometry at hand. 
The choice of a particular geometry and material defines 
$\tens{G}$. For instance, choosing vacuum recovers Eq.~\eqref{J0MainText}, as demonstrated in the Supplementary Material (SM), 
offering flexible design of dipole-dipole interactions through Eq.~\eqref{JijGeneral},
even for dipoles with fixed relative orientation. 

 Here we showcase the power of this approach through a particular example:  engineered dipole-dipole interactions enable a  `guide-slide' superabsorber that achieves enhanced absorption courtesy of a carefully calibrated interplay between its vibrational and optical environments~\cite{Brown2019}. 
This crucially relies on  structuring the energy levels due to Lamb-shifts induced by a dielectric metal sphere placed in the centre of a molecular dipole ring (see Fig.~\ref{fig:GuideSlideSchematic})
Achieving this without  engineered interactions necessitates sub-optimal alignment of dipoles (with reduced collective dipole moment---counterproductive to superabsorption) in addition to ad-hoc re-initialisation, and  embedding in a photonic band gap~\cite{Brown2019}. As we presently argue, engineering of the dipole-dipole interaction eliminates the need for such external control, enabling the design of simpler yet more powerful photonic devices, exemplifying the importance of the approach developed.

\begin{figure}
\includegraphics[width = 0.85\columnwidth]{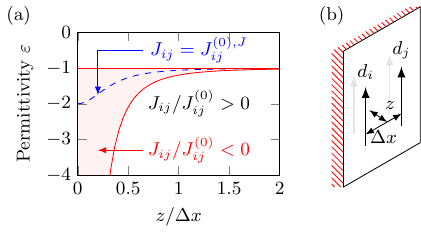}
    \caption{(a) Sign of the interaction matrix element $J_{ij}$ as the permittivity and dipole positions are varied near a dielectric surface as indicated in (b). The shaded region indicates where the sign of the interaction matrix element is flipped relative to the same aggregate in free space (matrix element $J_{ij}^{(0)}$).
    The blue dashed line shows where the interaction matrix element equals that of 
    a J-type aggregate (dipoles placed end-to-end) in free space with the same separation $\Delta x$. }
    \label{fig:PlanarParams}
\end{figure}

We begin by illustrating the engineering of dipolar interactions with
the minimal system for analytically demonstrating 
the tunable conversion of a physical H aggregate into an effective J aggregate. This opens up new avenues of research in biophysics and photophysics that depend on dipole-dipole interactions --- in our case enabling the design of a new type of light-harvesting superabsorber. We consider two dipoles in vacuum in the region $z>0$, with the region $z<0$ filled with a material of relative permittivity $\varepsilon$ (and unit relative permeability), in the electrostatic limit where any separations are assumed much shorter than the (reduced) transition wavelength $\lambda/2\pi$.  \cite{palacinoTuningCollectiveDecay2017,hemmerichInfluenceRetardationDielectric2018}) . Substituting the relevant $\mathbb{G}$ (\cite{palacinoTuningCollectiveDecay2017,hemmerichInfluenceRetardationDielectric2018}) into Eq.~\eqref{JijGeneral} produces an analytic result for the ratio of the interaction matrix element near a planar surface to its free space value ${J^{(0)}_{ij}}$ (see SM). Specialising to  dipoles at the same distance $z$ from the half-space, both with dipole moments oriented in the $y$ direction and separated in the $x$ plane by a distance $\Delta x $ ( see  Fig.~\ref{fig:PlanarParams}), one finds
$
    {J_{ij}}/{J^{(0)}_{ij}} = 1 - \frac{\varepsilon-1}{\varepsilon+1}[(2z/\Delta x)^2+1]^{-3/2}
$
(see SM). Since we seek cases where the interaction matrix element changes sign relative to free space, we require combinations of $z/\Delta x$ and $\varepsilon$ for which $ {J_{ij}}/{J^{(0)}_{ij}}<0$, achievable for $\varepsilon < -1$ (implying that a surface plasmon is the crucial ingredient) and distance to wall $z$ sufficiently shorter than pair separation $\Delta x$, as shown in Fig.~\ref{fig:PlanarParams}. Under these conditions,
an H-aggregate close to a surface supporting a surface plasmon assumes the interaction matrix element of a J-aggregate, as if its dipoles had been rotated (see SM for an interpretation via image dipoles). This is a remarkable result that shows for the first time the power of engineering dipole-dipole interactions for improving light harvesting efficiency. This has a wide-reaching impact on the study and development of quantum energy transport and storage systems that rely heavily on such dipolar couplings. 

Building on this simple yet remarkable observation for the dimer, we proceed to a ring of $N \geq 3$ dipoles, where a changed coupling sign is key to supporting superabsorbing behaviour \cite{Brown2019}. Previously, this was accomplished in a ring arrangement with dipole moments tilted appropriately with respect to each other.
By contrast, engineering the dipolar interactions  allows us to maximise the collective far-field absorption strength, while achieving the requisite energetic ordering through negative couplings with the help of a metallic sphere inside the ring as illustrated in Fig.~\ref{fig:GuideSlideSchematic}.  

 We deﬁne a guide-slide superabsorber to be a collection of optical dipoles with properties: (1) a ladder of excitation manifolds of a fixed number of dipole excitations, each with rapid relaxation to a well-deﬁned lowest-energy state within the manifold, (2) enhanced optical rates due to collective effects between dipoles coupling the lowest-energy states of adjacent manifolds~\cite{Brown2019}. 
To show the efficacy of our surface-plasmon-enabled guide-slide system we will utilise an open quantum systems heat engine framework and compare our new design against the previously proposed tilted guide-slide antenna~\cite{Brown2019}. 

We consider a ring assembly of interacting chromophores, each approximated as a two-level optical dipole. All dipoles couple to the common electromagnetic field and to individual vibrational baths, yielding the total Hamiltonian
\begin{equation}
    H = H_S + H_{\text{opt}} + H_{I,\text{opt}} + H_\text{vib} + H_{I,\text{vib}}, 
\end{equation}
where the Coulomb gauge system Hamiltonian describes the energies associated with just the dipoles \cite{Rouse_2019,Stokes_2018}
\begin{equation}
    H_S = \omega_0\sum_{i=1}^N \sigma_+^{(i)}\sigma_-^{(i)}+ \sum_{i\neq j=1, }^N J_{ij}\sigma_+^{(i)}\sigma_-^{(j)}
    \label{eq:sysham}
\end{equation}
with bare transition energies $\omega_0$ and  $J_{ij}$'s according to Eq.~\eqref{JijGeneral}; the $\sigma_\pm^{(n)}$ operators are the raising and lowering operators for the $n$th  dipole.
The free optical field obeys
$
    H_\text{opt} = \sum_p \omega_p a_p^\dag a_p,
$
with $\omega_p$ and $a^{(\dag)}_p$, representing the energy and annihilation (creation) operator, respectively, associated with the $p$th electromagnetic field mode. Dipole-electromagnetic interactions are captured by  
$
    H_{I,\text{opt}} = \sum_{i=1}^N  \sigma_x^{(i)} \sum_p \mathbf{d}_i \cdot \mathbf{e}_p f_p(a_p+a_p^\dag),
$
where $f_p$ determines the coupling strength of the $p$th photon mode (polarisation vector $\mathbf{e}_p$) to the dipole with vector dipole $\mathbf{d}_i$. Finally, each dipole couples to its own vibrational environment, with interaction Hamiltonian 
$
    H_{I,\text{vib}} = \sum_{i=1}^N \sigma_z^{(i)} \sum_v g_{i,v}(b_{i,v} + b^\dag_{i,v})
$
and
$
    H_{\text{vib }} = \sum_{i=1}^N \sum_v \omega_{i,v}b_{i,v}^\dag b_{i,v},
$
where $\omega_{i,v}$, $g_{i,v}$, and $b_{i,v}^{(\dag)}$ represent, respectively, the energy, coupling strength, and annihilation (creation) operator of mode $v$ at the $i$th dipole. Geometrically, we fix a nearest neighbour dipole separation of 2.5~nm inspired by natural molecular chromophore separations~\cite{Alo2023,Haver2019} and leaving space for a central metallic nanoparticle .  Slowly varying phases across the ring render the dipoles  approximately spatially indistinguishable by the electromagnetic field for relevant optical frequencies ($\approx700/2\pi\,$nm$\gg2.5$~nm). We let the radius of the inserted sphere be 1~nm shorter than the ring radius.

\begin{figure}
    \centering
    \begin{overpic}[width=\linewidth]{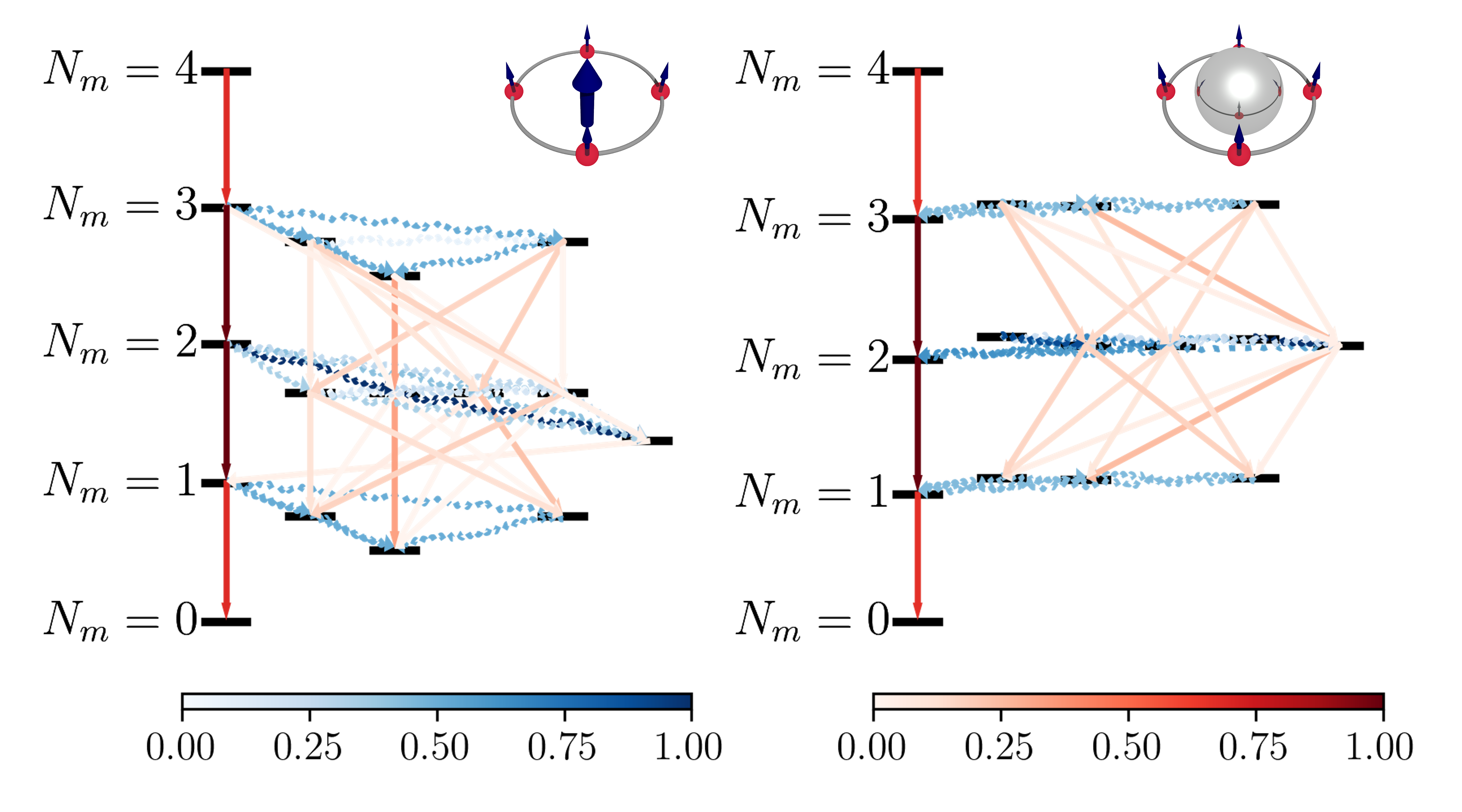}
     \put (18,50) {(a)}
 \put (65,50) {(b)}

    \end{overpic}
    \caption{Level transition schematic of a four-dipole ring (a) in free space (b) as spherical guide-slide superabsorber. Blue dotted arrows represent fast phonon relaxation inside manifolds of fixed excitation number, red arrows represent optical transitions between manifolds with manifold number $N_m$. Arrow saturation is scaled by the relative strength of the transition. Intra-manifold energy splitting is accentuated relative to manifold spacings. These are calculated by diagonalising $H_S$ and calculating transition rates from Fermi's golden rule.
    }
    \label{fig:LevelDiagram}
\end{figure}

The  eigenstructure of our system Hamiltonian $H_S$ \eqref{eq:sysham}  determines how optical and vibrational baths give rise to dissipative energy flow and conversion within an open quantum systems framework~\cite{Creatore2013,Killoran:2015aa, Fruchtman2016,Higgins:2017aa}. We expect maximal interaction with the optical field for dipoles aligned perpendicular to the ring plane, as individual dipoles become indistinguishable and form an effective super-dipole. Without the sphere, such a configuration in free space would result in an H-aggregate system with positive-valued $J_{ij}$ coupling elements. H-aggregates feature optically bright states at higher energy than optically dark states. Vibrational relaxation would result in rapid population transfer from the bright to the dark states, effectively trapping excitations in optically inactive states (see Fig.~\ref{fig:LevelDiagram}a), preventing superradiance and superabsorption. 

To mitigate against such phonon-induced breakdown of superabsorption, a tilted guide-slide configuration was proposed~\cite{Brown2019}, wherein each of the dipoles is tilted sufficiently from the azimuth to achieve a negative nearest neighbour interaction (in turn positioning dark states energetically above bright states). Then vibrational relaxation funnels excitations onto the Dicke ladder. However, the imperfect alignment of the dipoles also reduces the collectively enhanced dipole perpendicular to the ring plane, shifting some transition oscillator strength into destructive interference in the plane. 

Strategic engineering of nearest neighbour coupling, by the metallic sphere placed within the dipole ring, generates both negative nearest neighbour couplings and perfectly aligned dipoles in concert. This enables a guide-sliding effect due to vibrational relaxation alongside optimally enhanced optical rates courtesy of fully aligned dipoles. 
 Fig.~\ref{fig:LevelDiagram}b shows that
the most optically active states lie at the bottom of each manifold. With vibrational relaxation occurring much faster than optical decay, this setup fulfils our guide-slide conditions outlined above. 

We model the dynamics of our  system with the Bloch-Redfield master equation~\cite{TheoryOQSBook}  generating non-unitary dynamics for the reduced density matrix of the dipole ring due to dissipative influences from  vibrational and photon environments. The approximations implicit within this model are weak coupling for the system-bath coupling and that the bath relaxation timescale is much faster than the dynamics associated with the system, such that  bath correlations can be considered as delta functions, see SM for more details. This results in a master equation of the form
$\dot{\rho}_S(t) = -i[H_S,\rho_S]+(\mathcal{D}_\text{opt}+\mathcal{D}_\text{vib}) \rho_S$.
\
The optical dissipator $\mathcal{D}_{\text{opt}}$ is responsible for transitions between excitation number manifolds, through which sunlight photons pump our superabsorber. Conversely, $\mathcal{D}_\text{vib}$, leads to vibrational relaxation within excitation manifolds, and, dependent on eigenstructuring, manifesting as either guide-sliding or dark-state trapping. 

To quantify the efficiency of the guide-slide superabsorber, we employ a trap model for extracting and converting ring excitons to useful energy~\cite{Scully2011,Dorfman2013,Rahav2012}. This is  achieved by incoherently coupling an additional two-level system to the ring as an abstract load~\cite{Scully2011,Killoran:2015aa} (see SM).
Physically,  the trap could be  another dipole displaced below the centre of the ring with coupling to the ring dominated by  second-order photon field effects , i.e.~as a FRET process. From this trap dipole, excitations can then be  converted to useful energy.


Combining high-temperature solar photons with room-temperature phonons and the trap, we have thus constructed an effective quantum heat engine (QHE), allowing us to determine the power output of our ring antennas, and uncover how this power scales with the number of dipoles. We utilise a standard approach for calculating the power outputs of such a QHE \cite{Scully2011} by optimising the trap's power with respect to its decay rate that represents the load of the device (see SM for full details).
%

\begin{figure}
    
    \centering
    \begin{overpic}[width=\linewidth]{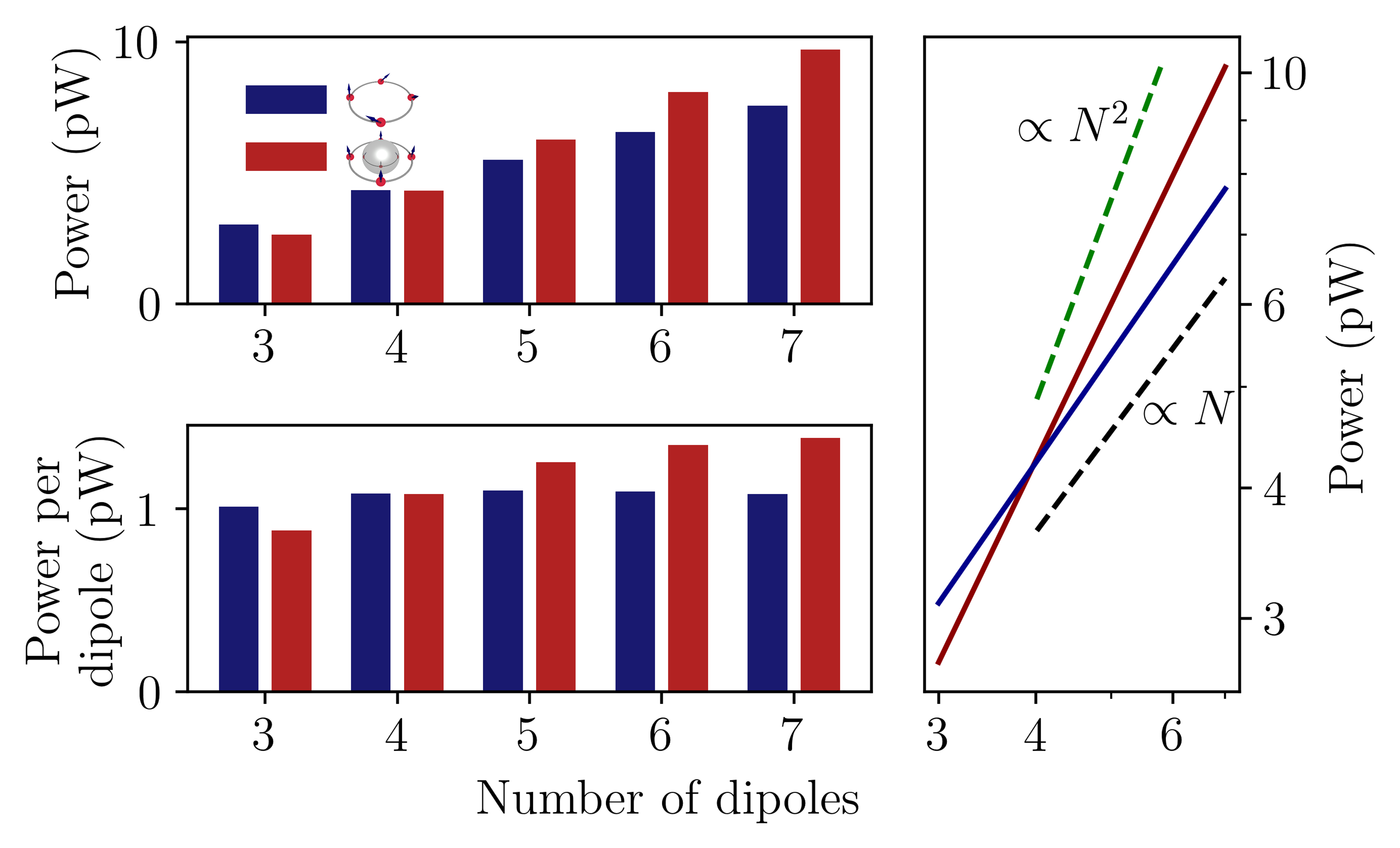}
         \put (32,54) {(a)}
 \put (70,54) {(b)}
 \put (32,27) {(c)}
    \end{overpic}
        \caption{Scaling of the solar cell power output for the configuration of the parallel spherical ring guide-slide, and the previously developed tilted guide-slide. The (a) power output (b) log-log plot of the power output (c) power output per dipole of the ring for increasing number of dipoles in the ring. The theoretical maximum Dicke scaling of $N^2$ (green dashed)
     and linear scaling (black dashed) are also plotted for comparison in (c). Full simulation parameters are given in the SM.}
    \label{fig:NScaling}
\end{figure}

We directly compare the power outputs of the solar absorbers using the tilted guide-slide developed previously and the novel surface plasmon-enabled (parallel) guide-slide developed here. Fig.~\ref{fig:NScaling} shows the results of numerical simulations for varying numbers of dipoles in the rings, for a permittivity of $\varepsilon = -2.37$, similar to that of chromium at $1.8$eV. Fig.~\ref{fig:NScaling}c depicts the power output per dipole, and we note an increase with $N$ for the spherical guide-slide setup, showing super-extensive scaling. We see for small values of $N$ the power output achieved by the tilted guide-slide is greater than that of the parallel guide-slide. This is due to the relative size of the sphere to the ring radius being small in this limit, restricting the ability to engineer dipolar interactions. By contrast, increasing ring radius improves
the approximation of dipoles experiencing a planar dielectric interface, meaning the design is scalable to larger $N$.
In Fig.~\ref{fig:NScaling}b, a log-log plot shows the exponent for each configuration's power output of the form $P=\alpha N^m$. 
Our surface plasmon-enabled guide-slide has greater super-extensive power scaling than the tilted guide-slide, with a growth exponent $m = 1.55$ vs.~$m=1.08$. These growth exponents were calculated using linear regression with sum squared residuals of $0.0049$ and $0.0021$ respectively.  This increase in super-linear scaling is due to the parallel dipoles in the surface plasmon-enabled guide-slide offering maximal collective enhancement. 


Our results exemplify the  utility of selectively designing the  electromagnetic field through metals or dielectrics to achieve bespoke couplings different from far-field behaviour. While incorporating a spherical metallic nano-particle into the centre of a nanoscopic molecular ring of dipoles presents an implementation challenge, recent experimental studies have validated the approach of utilising DNA origami to interface molecular rings with nano-particles ~\cite{DNAOrigami1,DNAOrigami2,DNAOrigami3}, supporting the in-principle feasibility of our outlined design. Its simplicity, enhanced super-linear scaling, and operability without a photonically engineered environment and active re-initialisation establish our present design as a leading candidate for a quantum-enhanced molecular antenna. Structuring the local electromagnetic field modes to induce the engineered dipole-dipole interactions also impacts the optical decay rates and yields a secondary advantage (see SM). Specifically, we observe an additional enhancement of the super-linear behaviour from a $N$-dependent Purcell effect. As this effect is not directly related to realising superabsorption, as the counterpart to Dicke superradiance, to avoid ambiguity we have opted not to include this effect in this discussion. While our design---unlike previous superabsorbing ring proposals---performs well without a photonic band-gap and active reinitisalition, it benefits and maintains its advantage when available leading to greater super-extensive scaling, approaching the theoretical limit (see SM). Taking into account the effects of dephasing, detuning and absorption into the dielectric we also determined that the surface plasmon-enabled guide-slide is remarkably robust all these types of deleterious effects even at moderate levels (see SM). 

In summary, we have demonstrated that surface plasmons allow the engineering of molecular dipole-dipole interactions, thereby enabling advantageous couplings for light-harvesting systems. Specifically, we have shown the possibility of combining desirable far-field H-aggregate-like properties with effective J-aggregate-like near-field couplings. Building upon this, we have developed a novel superabsorber by precisely engineering the Hamiltonian of a molecular dipole ring through the strategic placement of a metallic nanoparticle at its center. Within the framework of a quantum heat engine, we have shown that our design surpasses previous proposals in performance, and, due to its greater simplicity, represents an ideal candidate for the development of light-harvesting superabsorbers. Importantly, we have illustrated that the innovative design of the local dielectric or metallic environment enables the engineering of system Hamiltonians with properties conducive to the design and implementation of quantum-enhanced devices, representing a critical advancement towards realising efficient energy transport systems. The paradigm introduced offers profound and wide-reaching impact across a broad range of quantum technologies which rely on both engineered interactions and coupling to an external environment.  

\subsection*{Acknowledgments} The authors thank Will Brown for helpful discussions. A. Burgess and E.M. Gauger thank the Leverhulme Trust for support through grant number RPG-2022-335. M.C.~Waller thanks the EPSRC Doctoral Training Programme EPSRC/DTP 2020/21/EP/T517896/1, and R. Bennett acknowledges financial support from EPSRC grant EP/W016486/1. 

%

\newpage
\widetext
\begin{center}
\textbf{\large Engineering dipole-dipole couplings for enhanced cooperative light-matter interactions: Supplementary Material}
\end{center}

\setcounter{equation}{0}
\setcounter{figure}{0}
\setcounter{table}{0}
\setcounter{page}{1}
\makeatletter
\renewcommand{\theequation}{S\arabic{equation}}
\renewcommand{\thefigure}{S\arabic{figure}}
\renewcommand{\bibnumfmt}[1]{[S#1]}
\renewcommand{\citenumfont}[1]{S#1}

\input{Supplementary}

\end{document}

%% file: Supplementary.tex
\section{Green's tensors}

The Green's tensor $\tens{G}(\vec{r},\vec{r}',\omega)$ that solves the Helmholtz equation for an inhomogeneous region is most usefully written as the sum of two parts;
\begin{equation}
    \tens{G}(\vec{r},\vec{r}',\omega) = \tens{G}^{(0)}(\vec{r},\vec{r}',\omega) + \tens{G}^{(1)}(\vec{r},\vec{r}',\omega),
\end{equation}
where $\tens{G}^{(0)}(\vec{r},\vec{r}',\omega)$ is the homogenous `background' part, and $\tens{G}^{(1)}(\vec{r},\vec{r}',\omega)$ (known as the `scattering' Green's tensor) contains all the information about any boundaries between regions of different permittivities and/or permeabilities. In both the cases considered here, the background Green's tensor will be that of vacuum, since the Green's function is only evaluated at the positions of the dipoles, which are all in vacuum. The vacuum Green's tensor for distinct source and observation positions is given by (see, e.g. Appendix B of \cite{buhmann2012Book1});
\begin{equation}\label{fullVacuumG}
    \tens{G}^{(0)}(\vec{r},\vec{r}',\omega) = -\frac{e^{i \omega\rho}}{4\pi \omega^2 \rho^3} \left\{\left[ 1 - i \omega \rho - (\omega \rho)^2\right]\mathbb{I} - \left[3 - 3i \omega \rho - (\omega \rho)^2 \right] \mathbf{e}_\rho \otimes  \mathbf{e}_\rho  \right\},
\end{equation}
where $\mathbf{e}_\rho$ is a unit vector in the direction of $\bm{\rho} = \vec{r}-\vec{r}'$, and $\rho = |\bm{\rho}|$. As in the main text, we have used units where $c=1$, so that for example the wavelength $\lambda$ is given by $\lambda = 2\pi/\omega$. In the electrostatic limit (defined by $ \omega \rho \ll 1$, or equivalently $\rho \ll \lambda/(2\pi)$), only the first terms in each of the square brackets in \eqref{fullVacuumG} survive, giving
\begin{equation}\label{G0Electrostatic}
    \tens{G}^{(0)}_{\omega\rho \ll 1}(\vec{r},\vec{r}',\omega) = -\frac{c^2}{4 \pi \omega^2 \rho^3} (\mathbb{I} - 3 \mathbf{e}_\rho \otimes \mathbf{e}_\rho)
\end{equation}
as the electrostatic vacuum Green's tensor. 

\subsection{Dielectric half-space, electrostatic limit}

The Green's tensor $\tens{G}^\text{HS}_{\omega\rho \ll 1}$ for a dielectric half-space in the electrostatic limit is given by \cite{palacinoTuningCollectiveDecay2017,hemmerichInfluenceRetardationDielectric2018}:
\begin{align}
    \tens{G}^\text{HS}_{\omega\rho \ll 1}&(\vec{r},\vec{r}',\omega) = \tens{G}^{(0)}_{\omega\rho \ll 1}(\vec{r},\vec{r}',\omega) + \frac{\varepsilon-1}{\varepsilon+1} \tens{G}^{(0)}_{\omega\rho \ll 1}(\vec{r},\bar{\vec{r}}',\omega)\cdot \mathrm{diag}(-1,-1,1) = \tens{G}^{(0)}_{\omega\rho \ll 1}(\vec{r},\vec{r}',\omega) + \tens{G}^{(1)}_{\omega\rho \ll 1}(\vec{r},\vec{r}',\omega)\label{GPlane},
\end{align}
where $\tens{G}^{(0)}_{\omega\rho \ll 1}(\vec{r},\vec{r}',\omega)$ is given by Eq.~\eqref{G0Electrostatic}, and $\bar{\vec{r}} = (x,y,-z)$, as familiar from an electrostatic image construction. The second term is the scattering part of the Green's tensor. The interaction matrix element
\begin{equation}
    J_{ij} = -{\omega^2} \vec{d}_i \cdot \mathrm{Re}[ \tens{G}(\vec{r}_i,\vec{r}_j,\omega)] \cdot \vec{d}_j = J_{ij}^{(0)} + J_{ij}^{(1)}
\end{equation}
can therefore be straightforwardly calculated from this by substituting in the electrostatic limit of the vacuum Green's tensor. The free space contribution $J_{ij}^{(0)}$ to $J_{ij}$ is therefore given by;
\begin{equation}
    J_{ij}^{(0)}= -\omega^2 \vec{d}_i \cdot \mathrm{Re}[ \tens{G}^{(0)}_{\omega\rho \ll 1}(\vec{r}_i,\vec{r}_j,\omega)] \cdot \vec{d}_j = \frac{1}{4\pi \rho^3} \vec{d}_i \cdot (\mathbb{I} - 3 \mathbf{e}_\rho \otimes \mathbf{e}_\rho) \cdot \vec{d}_j = \frac{1}{4\pi \rho^3}  [\vec{d}_i \cdot \vec{d}_j- 3 (\vec{d}_i \cdot \mathbf{e}_\rho ) (\vec{d}_j \cdot \mathbf{e}_\rho) ]\label{J0}
\end{equation}
as is well-known. The configuration of interest in the main text has $\vec{d}_i = \vec{d}_j = (0,d,0)$, and a separation vector of length $\Delta x$ along the $x$ direction. In this case, the free-space coupling matrix element becomes;
\begin{equation}
    J_{ij}^{(0)} = \frac{d^2}{4\pi\varepsilon_0 \Delta x^3}  .
\end{equation}
To work out the additional part $J^{(1)}_{ij}$ that is dependent on the surface (or, in other words, the part that is given by the scattering Green's tensor $\tens{G}^{(1)}_{\omega\rho \ll 1}$, it is simpler to note that our chosen dipole orientation means we only require the $yy$ component of $\tens{G}^{(1)}_{\omega\rho \ll 1} = \frac{\varepsilon-1}{\varepsilon+1} \tens{G}^{(0)}(\vec{r},\bar{\vec{r}}',\omega)\cdot \mathrm{diag}(-1,-1,1)$, given by:
\begin{equation}
   -\frac{\varepsilon-1}{\varepsilon+1} \tens{G}^{(0)}_{\omega\rho \ll 1,yy}(\vec{r},\bar{\vec{r}}',\omega) = \frac{\varepsilon-1}{\varepsilon+1} \frac{c^2}{4 \pi \omega^2 \bar{\rho}^3} [1 - 3 (\mathbf{e}_{\bar{\rho}} \otimes \mathbf{e}_{\bar{\rho}})_{yy}] = \frac{\varepsilon-1}{\varepsilon+1} \frac{c^2}{4 \pi \omega^2 \bar{\rho}^3}
\end{equation},
where $\bar{\rho} = |\vec{r}-\bar{\vec{r}}'|$. This gives  
\begin{equation}
  J^{(1)}_{ij}= - \frac{\varepsilon-1}{\varepsilon+1} \frac{1}{4 \pi \varepsilon_0}\frac{d^2 }{ (\Delta x^2 + 4z^2)^{3/2}}.
\end{equation}
The ratio between the coupling matrix element in the absence or presence of the surface is therefore given by;
\begin{equation}\label{JRatioSupp}
    \frac{J_{ij}}{J^{(0)}_{ij}} = \frac{J^{(0)}_{ij} + J^{(1)}_{ij}}{J^{(0)}_{ij}} = 1+ \frac{\Delta J_{ij}}{J^{(0)}_{ij}} = 1   - \frac{\varepsilon-1}{\varepsilon+1} \frac{\Delta x^3}{(\Delta x^2 + 4z^2)^{3/2}} =  1   - \frac{\varepsilon-1}{\varepsilon+1} \frac{1}{(1 + 4z^2/\Delta x^2)^{3/2}}
\end{equation}
as in the main text. We are interested in when $\frac{J_{ij}}{J^{(0)}_{ij}}$ changes sign -- this happens at two places. The first is the limit $\varepsilon \to -1$, where the ratio diverges in opposite directions either side. The other is found by solving $\frac{J_{ij}}{J^{(0)}_{ij}}=0$ for $\varepsilon$, which gives:
\begin{equation}
    \varepsilon = \frac{1+ (1 + 4z^2/\Delta x^2)^{3/2}}{1- (1 - 4z^2/\Delta x^2)^{3/2}}.
\end{equation}
These two limits represent the upper and lower bounds, respectively, of the shaded region shown in the figure in the main text.   

We also require the interaction matrix element for the equivalent J-aggregate in free space, which is the same except with $\vec{d}_i = \vec{d}_j = (d,0,0)$. In this case Eq.~\eqref{J0} becomes;
\begin{equation}
  J^{(0),J}_{ij} =  \frac{1}{4\pi\varepsilon_0 \Delta x^3}  [d^2- 3 d^2  ] = -\frac{d^2}{2\pi\varepsilon_0 \Delta x^3}
\end{equation}
and the ratio to the (H-aggregate) coupling matrix element $J_{ij}$ that includes the presence of the surface becomes;
\begin{equation}
    \frac{J_{ij}}{J^{(0),J}_{ij}}
    = \frac{\frac{1}{\Delta x^3}- \frac{\varepsilon-1}{\varepsilon+1}\frac{1}{(\Delta x^2 + 4z^2)^{3/2}}}{-\frac{2}{\Delta x^3}} = \frac{1}{2}\left[\frac{\varepsilon-1}{\varepsilon+1}\frac{1}{(1 + 4z^2/\Delta x^2)^{3/2}} - 1\right].
\end{equation}
Solution of the equation ${J_{ij}}/{J^{(0),J}_{ij}} = 1$ for $\varepsilon$ yields;
\begin{equation}
   \varepsilon =  \frac{1+(3+12(z/\Delta x)^2) \sqrt{4 (z/\Delta x)^2+1}}{1-(3+12(z/\Delta x)^2) \sqrt{4 (z/\Delta x)^2+1}}
\end{equation}
which is the equation of the dashed line in the figure in the main text. It is also possible to deduce the flipped sign of $J_{ij}$ in the near-field by an image construction, as shown in Fig.~\ref{imageConstruction}. 

\begin{figure}
 \begin{tikzpicture}

 \def\hWidth{2}
 \def\vHeight{3}
 \def\angle{30}
 \def\thickness{0.2}
 \def\dipoleSep{01}
 \def\dipoleLength{1}
 \def\dipolePosX{0.3}
 \def\dipolePosY{0.4}
 \def\reflectionOffset{0.2}
 \def\shift{4.5}
 
 
       \node at (-0.5,3.5) {$b)$};
 \draw (0,0) --++ (0,\vHeight) -- ++({\hWidth*cos(\angle)},{\hWidth*sin(\angle)}) --++ (0,-\vHeight) -- cycle;
 \draw[->,thick] (\dipolePosX*\hWidth,\dipolePosY*\vHeight) -- ++(0,\dipoleLength) node[anchor = south] {$d_i$};
 \draw[->,thick] ({\dipolePosX*\hWidth+\dipoleSep*cos(\angle)},{\dipolePosY*\vHeight+\dipoleSep*sin(\angle)}) -- ++(0,\dipoleLength) node[anchor = south] {$d_j$};
 \draw[<->] (\dipolePosX*\hWidth,\dipolePosY*\vHeight+0.2) -- ++({\dipoleSep*cos(\angle)},{\dipoleSep*sin(\angle)}) node[anchor = north,midway] {$\Delta x$} ;
\draw[dotted] (0,0.5*\vHeight) --++ ({\hWidth*cos(\angle)},{\hWidth*sin(\angle)});
\draw[dotted]  ({0.1*\hWidth*cos(\angle)},{0.1*\hWidth*sin(\angle)}) --++(0,\vHeight);
       \node at (1.62,1.08) {$\varepsilon $};
 
  \draw[<->] ({\dipolePosX*\hWidth + 0*\dipoleSep*cos(\angle)},{\dipolePosY*\vHeight+0.2 + 0*\dipoleSep*sin(\angle)}) -- ++(-0.4,0.2) node[anchor = north,midway] {$z$} ;
   \node at (3,2.2) {$=$};
  
 \tikzset{shift={(\shift,0)}}  
 

 \draw[->,thick] (\dipolePosX*\hWidth,\dipolePosY*\vHeight) -- ++(0,\dipoleLength) node[anchor = south] {$d_i$};
  \draw[<-,thick] (\dipolePosX*\hWidth -0.8,\dipolePosY*\vHeight+0.4) -- ++(0,\dipoleLength) node[anchor = south] {$ \frac{\varepsilon - 1}{\varepsilon + 1}d_i$};
 \draw[->,thick] ({\dipolePosX*\hWidth+\dipoleSep*cos(\angle)},{\dipolePosY*\vHeight+\dipoleSep*sin(\angle)}) -- ++(0,\dipoleLength) node[anchor = south] {$d_j$};

 \draw[<->] (\dipolePosX*\hWidth,\dipolePosY*\vHeight+0.2) -- ++({\dipoleSep*cos(\angle)},{\dipoleSep*sin(\angle)}) node[anchor = north,midway] {$\Delta x$} ;
\draw[dotted] (0,0.5*\vHeight) --++ ({\hWidth*cos(\angle)},{\hWidth*sin(\angle)});
\draw[dotted]  ({0.1*\hWidth*cos(\angle)},{0.3*\vHeight + 0.1*\hWidth*sin(\angle)}) --++(0,0.5*\vHeight);
 
  \draw[<->] ({\dipolePosX*\hWidth + 0*\dipoleSep*cos(\angle)},{\dipolePosY*\vHeight+0.5 + 0*\dipoleSep*sin(\angle)}) -- ++(-0.8,0.4) node[anchor = south,midway] {$2z$} ;

     \node at (3,2.2) {$\approx$};
          \node at (3,1.7) {\scriptsize$\left(\frac{\varepsilon - 1}{\varepsilon + 1} \gg 1\right)$};
     
\tikzset{shift={(\shift,0)}}  


\draw[<-,thick] (\dipolePosX*\hWidth -0.8,\dipolePosY*\vHeight+0.4) -- ++(0,\dipoleLength) node[anchor = south] {$ \frac{\varepsilon - 1}{\varepsilon + 1}d_i$};
\draw[->,thick] ({\dipolePosX*\hWidth+\dipoleSep*cos(\angle)},{\dipolePosY*\vHeight+\dipoleSep*sin(\angle)}) -- ++(0,\dipoleLength) node[anchor = south] {$d_j$};

\draw[<->] (\dipolePosX*\hWidth -0.8,\dipolePosY*\vHeight+0.4+0.5*\dipoleLength+0.2) --++ (1.65,0) node[anchor = north,midway] {\tiny $\sqrt{\Delta x^2 + (2z)^2}$};

\tikzset{shift={(-3*\shift,0)}}  


 \def\shortDipole{0.6}
  \def\verticalSpace{1}
    \def\horizontalSpace{1}
    \def\origin{0.5}
    
       \node at (-1,3.5) {$a)$};
\fill[white] (-0.2,0) rectangle (0,0.3);

\node[anchor = west] at (2,2.8) {$J_{ij}>0$};
\draw[rounded corners] (-1.5*\shortDipole, \origin-0.1) rectangle ++(2.8*\horizontalSpace, 0.8);
\draw[->,thick] (-\shortDipole*0.5,\origin+0.5*\shortDipole) -- ++(\shortDipole,0);
\draw[->,thick] (\horizontalSpace-\shortDipole*0.5,\origin+0.5*\shortDipole) -- ++(\shortDipole,0);

\draw[rounded corners] (-1.5*\shortDipole, \origin+\verticalSpace-0.1) rectangle ++(2.8*\horizontalSpace, 0.8);
\draw[->,thick] (0,\origin+\verticalSpace) -- ++(0,\shortDipole);
\draw[<-,thick] (\horizontalSpace,\origin+\verticalSpace) -- ++(0,\shortDipole);

\draw[rounded corners] (-1.5*\shortDipole, \origin+2*\verticalSpace-0.1) rectangle ++(2.8*\horizontalSpace, 0.8);
\draw[->,thick] (0,\origin+2*\verticalSpace) -- ++(0,\shortDipole);
\draw[->,thick] (\horizontalSpace,\origin+2*\verticalSpace) -- ++(0,\shortDipole);

\draw (1.8,0.3) --++ (0.2,0) -- ++(0,2)node[anchor = west,midway] {$J_{ij}<0$}  -- ++(-0.2,0);
\draw[dashed](3.5,-0.2) --++ (0,4);
 \end{tikzpicture}
 \caption{Image construction. In panel a) we note for reference the sign of $J_{ij}$ for three different mutual orientations of dipoles $i$ and $j$. In panel b) we place parallel dipoles (an H-aggregate) near a plane with permittivity $\varepsilon$ and construct the corresponding image charge problem (ignoring the self-interaction of dipole $j$ with its own image as it is irrelevant to the coupling to dipole $i$). If the quantity $\frac{\varepsilon-1}{\varepsilon-1}$ can be made sufficiently large (which happens for a given geometry when $\varepsilon$ is negative, but not too negative, as shown in Fig.~1 in the main text) the contribution of the image dipole to the interaction will drown out that of the original dipole $d_i$. This leads the system to behave as if it were in the arrangement shown on the right hand side, where the distance between the dipole $i$ and the image of $j$ appears in the denominator of Eq.~\eqref{JRatioSupp}, as expected. This antiparallel orientation of dipoles has $J_{ij}<0$, which is the feature required for guide-slide behavior, and is common to the physically important head-to-tail arrangement (a J-aggregate) discussed in the main text.}\label{imageConstruction}
\end{figure}

\subsection{Sphere}
Here we give the Green's tensor for two bodies near a magnetoelectric sphere of relative permittivity $\varepsilon(\omega)$, relative permeability $\mu(\omega)$, and radius $R$. In contrast with our simple example above, we will consider all general distances, not just the electrostatic limit. Just as with the half-space, the Green's tensor of the sphere is decomposed into 
\begin{align}
    \tens{G}_\mathrm{sphere}&(\vec{r},\vec{r}',\omega) = \tens{G}^{(0)}(\vec{r},\vec{r}',\omega) +\tens{G}^{(1)}_\mathrm{sphere}(\mathbf{r},\mathbf{r}',\omega),
\end{align}
We use the spherical coordinate system $\mathbf{r}=(r,\theta,\phi)$, where $r$ is the radial distance, $\theta$ is the polar angle, and $\phi$ is the azimuthal angle, and choose the centre of the spherical body to align with the origin of the coordinate system. The spherical variables are related to Cartesian coordinates via,
\begin{equation}
    x= r \sin\theta \cos\phi, \quad 
    y= r \sin\theta \sin\phi, \quad
    z= r \cos\theta.
\end{equation}
In this coordinate system, the scattering Green's tensor for the sphere is \cite{buhmann2012Book1};
\begin{align}
    &\tens{G}^{(1)}_{\mathrm{sphere}}(\mathbf{r},\mathbf{r}',\omega)= \frac{i\omega}{4\pi} \sum^\infty_{n=1} \frac{2n+1}{n(n+1)} \sum^n_{m=0} \frac{(n-m)!}{(n+m)!}(2-\delta_{0m}) \notag \\
    \times& \sum_{p=\pm 1} \left[ B^M_n \mathbf{M}_{nm,p}(\mathbf{r},\omega) \otimes \mathbf{M}_{nm,p}(\mathbf{r}',\omega) + B^N_n \mathbf{N}_{nm,p}(\mathbf{r},\omega) \otimes \mathbf{N}_{nm,p}(\mathbf{r}',\omega)
    \right],
    \label{eq:G_Sph_Initial}
\end{align}
where $\mathbf{M}_{nm,p}(\mathbf{r},q)$ and $\mathbf{N}_{nm,p}(\mathbf{r},q)$ are even $(p=+1)$ and odd $(p=-1)$ spherical wave vector functions, given explicitly by,
\begin{subequations}
    \begin{align}
        \mathbf{M}_{nm,-1}(\mathbf{r},q) =& \frac{m}{\sin \theta} h_n^{(1)}(qr) P^m_n(\cos \theta) 
        \cos (m\phi) \mathbf{e}_\theta -h_n^{(1)}(qr) \frac{dP^m_n(\cos\theta)}{d\theta} \sin(m\phi) \mathbf{e}_\phi \label{eq:Mnm_-1}, 
        \\
        \mathbf{M}_{nm,+1}(\mathbf{r},q) =& - \frac{m}{\sin \theta} h_n^{(1)}(qr) P^m_n(\cos \theta) 
        \sin (m\phi) \mathbf{e}_\theta -h_n^{(1)}(qr) \frac{dP^m_n(\cos\theta)}{d\theta} \cos(m\phi) \mathbf{e}_\phi \label{eq:Mnm_+1},
        \\
        \mathbf{N}_{nm,-1}(\mathbf{r},q) =& \frac{n(n+1)}{qr} h_n^{(1)}(qr) P^m_n(\cos \theta) 
        \sin (m\phi) \mathbf{e}_r \notag \\ 
        + \frac{1}{qr}& \frac{d\left[ qrh_n^{(1)}(qr)\right]}{dr} \left( \frac{dP^m_n(\cos\theta)}{d\theta} \sin(m\phi) \mathbf{e}_\theta + \frac{m}{\sin\theta} P^m_n(\cos\theta) \cos(m\phi) \mathbf{e}_\phi \right) \label{eq:Nnm_-1},
        \\
        \mathbf{N}_{nm,+1}(\mathbf{r},q) =& \frac{n(n+1)}{qr} h_n^{(1)}(qr) P^m_n(\cos \theta) 
        \cos (m\phi) \mathbf{e}_r \notag \\ 
        + \frac{1}{qr}& \frac{d\left[ qrh_n^{(1)}(qr)\right]}{dr} \left( \frac{dP^m_n(\cos\theta)}{d\theta} \cos(m\phi) \mathbf{e}_\theta - \frac{m}{\sin\theta} P^m_n(\cos\theta) \sin(m\phi) \mathbf{e}_\phi \right) \label{eq:Nnm_+1},
    \end{align}
\end{subequations}
where $H_n^{(1}(x)$ is the spherical Hankel function of the first kind, $P^m_n(x)$ are the Legendre functions and $\mathbf{e}_{r/\theta/\phi}$ are unit vectors in their respective directions. The coefficients $B^M_n$ and $B^N_n$ are the Mie reflection coefficients \cite{mie_beitrage_1908} given by,
\begin{subequations}
    \begin{align}
        B^M_n(\omega) &= \frac{\mu(\omega) \left[z J_l(z) \right]' J_l(z_1) - J_l(z)\left[ z_1 J_l(z_1)\right]'}{\mu(\omega) \left[z H_l(z) \right]' J_l(z_1) - H_l(z)\left[ z_1 J_l(z_1)\right]'}, 
        \\
        B^N_n(\omega) &= - \frac{\varepsilon(\omega) \left[z J_l(z) \right]' J_l(z_1) - J_l(z)\left[ z_1 J_l(z_1)\right]'}{\varepsilon(\omega) \left[z H_l(z) \right]' J_l(z_1) - H_l(z)\left[ z_1 J_l(z_1)\right]'},
    \end{align}
\end{subequations}
where $J_l(z)$ is the spherical Bessel function of the first kind, $z=\omega R$, $z_1=\sqrt{\varepsilon(\omega)\mu(\omega)} z$ and the prime denotes differentiation with respect to the respective argument. 

To put this into a usable form, we place the source and observation points $\mathbf{r}$ and $\mathbf{r}'$ on the $x-z$ plane and assume that they are on opposite sides of the $z$-axis, such that one position has $\phi_\mathrm{A}=0$ and the other has $\phi_\mathrm{B}=\pi$. Due to the spherical symmetry of the ring system considered in this paper, this assumption can be made for any two of the dipoles making up the ring around the sphere. This means we label the positions of two bodies as $\mathbf{r}_\mathrm{A}$ and $\mathbf{r}_\mathrm{B}$ 
defined in spherical coordinates as,
\begin{align}
    \mathbf{r}_\mathrm{A} &= (r_\mathrm{A}, \theta_\mathrm{A},0), &
    \mathbf{r}_\mathrm{B} &= (r_\mathrm{B}, \theta_\mathrm{B},\pi).
\end{align}
To perform the summations over $m$ in $\tens{G}^{(1)}_{\mathrm{sphere}}(\mathbf{r}_\mathrm{A},\mathbf{r}_\mathrm{B},\omega)$, we can make use of the addition theorem for spherical harmonics \cite{john_david_jackson_classical_1975},
\begin{equation}
    \sum^n_{m=0} C_{nm} P^m_n(\cos \theta)  P^m_n(\cos \theta') \cos(m\phi) = P_n(\psi),\label{eq:addition_theorem} 
\end{equation}
where 
\begin{equation}
    \psi = \cos \theta \cos \theta' + \sin \theta \sin \theta' \cos \phi .
\end{equation}
To apply \eqref{eq:addition_theorem} to our system, we make the substitutions $\theta \rightarrow \theta_\mathrm{A}$, $\theta' \rightarrow \theta_\mathrm{B}$ and $\phi \rightarrow \phi_\mathrm{B} = \pi$, producing,
\begin{equation}
    \sum^n_{m=0} C_{nm} P^m_n(\cos\theta_\mathrm{A})P^m_n(\cos\theta_\mathrm{B}) = P_n(\gamma) ,\label{eq:addition_theorem_applied}
\end{equation}
where we have defined $\Theta \equiv \theta_\mathrm{A}+\theta_\mathrm{B}$ and $\gamma \equiv \cos \Theta$. 
We can manipulate this identity into different forms to use in each of the terms of $\tens{G}^{(1)}(\mathbf{r}_\mathrm{A},\mathbf{r}_\mathrm{B},\omega)$. By differentiating \eqref{eq:addition_theorem} twice with respect to $\phi$ we obtain,
\begin{align}
    \sum^n_{m=0} m^2 C_{nm} P^m_n(\cos \theta)  P^m_n(\cos \theta') \cos(m\phi) 
    &= - \left[ \frac{d^2\psi}{d \phi^2} \frac{dP_n(\psi)}{d\psi} 
    +  \left(\frac{d\psi}{d \phi}\right)^2  \frac{d^2P_n(\psi)}{d\psi^2} \right] \notag 
    \\
    &= \sin \theta \sin \theta' \left( \cos \phi \frac{dP_n(\psi)}{d\psi} 
    + \sin \phi \frac{d^2P_n(\psi)}{d\psi^2} \right).
\end{align}
Applying to our system via the above substitutions results in,
\begin{align}
    \sum^n_{m=0} m^2 C_{nm} P^m_n(\cos\theta_\mathrm{A})P^m_n(\cos\theta_\mathrm{B}) 
    &= - \sin \theta_\mathrm{A} \sin \theta_\mathrm{B} P'_n(\gamma) .\label{eq:addition_theorem_1}
\end{align}
Similarly, differentiating \eqref{eq:addition_theorem} with respect to $\theta$ or $\theta'$ and applying to our system in the same way produces,
\begin{align}
    \sum^n_{m=0} C_{nm} P^m_n(\cos \theta_\mathrm{A}) \frac{dP^m_n(\cos \theta_\mathrm{B})}{d\theta_\mathrm{B}} \cos(m\pi) 
    = \sum^n_{m=0} C_{nm}  \frac{dP^m_n(\cos \theta_\mathrm{A})}{d\theta_\mathrm{A}} P^m_n(\cos \theta_\mathrm{B}) \cos(m\pi) = - \sin\Theta P'_n(\gamma). \label{eq:addition_theorem_2}
\end{align}
Finally we differentiate \eqref{eq:addition_theorem} with respect to $\theta$ and $\theta'$, giving us,
\begin{equation}
    \sum^n_{m=0} C_{nm} \frac{dP^m_n(\cos \theta)}{d\theta}  \frac{dP^m_n(\cos \theta)}{d\theta} \cos(m\phi) = \frac{d\psi}{d\theta'} \frac{d^2P_n(\psi)}{d\psi^2} \frac{d\psi}{d\theta} + \frac{dP_n(\psi)}{d\psi} \frac{d^2\psi}{d\theta' d\theta}.
\end{equation}
Finally, applying this to our system produces,
\begin{align}
    \sum^n_{m=0} C_{nm} \frac{dP^m_n(\cos \theta_\mathrm{A})}{d\theta_\mathrm{A}}  \frac{dP^m_n(\cos \theta_\mathrm{B})}{d\theta_\mathrm{B}} \cos(m\pi) = (1-\gamma^2) P''_n(\gamma) - \gamma P'_n(\gamma) 
    = \gamma P'_n(\gamma) - n(n+1) P_n(\gamma), \label{eq:addition_theorem_3}
\end{align}
where in the final step we have made use of the Legendre equation \cite{john_david_jackson_classical_1975},
\begin{equation}
    (1-x^2)\frac{d^2P}{dx^2} -2x \frac{dP}{dx} + n(n+1)P = 0 .
\end{equation}
Now we can substitute \eqref{eq:addition_theorem_applied}, \eqref{eq:addition_theorem_1}, \eqref{eq:addition_theorem_2} and \eqref{eq:addition_theorem_3} into $\tens{G}^{(1)}(\mathbf{r}_\mathrm{A},\mathbf{r}_\mathrm{B},\omega)$ to generate an expression for the spherical scattering Green's tensor,
\begin{equation}
    \tens{G}^{(1)}_\mathrm{sphere}(\mathbf{r}_\mathrm{A},\mathbf{r}_\mathrm{B},\omega) = \sum_{i,j=r,\theta,\phi} \tens{G}^{(1)}_{ij}(\mathbf{r}_\mathrm{A},\mathbf{r}_\mathrm{B},\omega) \,\mathbf{e}_{i_\mathrm{A}} \otimes \mathbf{e}_{j_\mathrm{B}},
\end{equation}
with the non-zero components given by,
\begin{subequations}\label{eq:G1_Sph}
    \begin{align}
        G_{rr}^{(1)}(\mathbf{r}_\mathrm{A},\mathbf{r}_\mathrm{B},\omega) &= \frac{i}{4\pi \omega r_\mathrm{A} r_\mathrm{B}} \sum^\infty_{n=1} n(n+1)(2n+1) B^N_n(\omega) P_n(\gamma) Q_n^{(1)},
        \\
        G_{r\theta}^{(1)}(\mathbf{r}_\mathrm{A},\mathbf{r}_\mathrm{B},\omega) &= -\frac{i \sin\Theta}{4\pi \omega r_\mathrm{A} r_\mathrm{B}} \sum^\infty_{n=1} (2n+1)B^N_n(\omega) P'_n(\gamma) Q_n^{(2)},
        \\
        G_{\theta r}^{(1)}(\mathbf{r}_\mathrm{A},\mathbf{r}_\mathrm{B},\omega) &= -\frac{i \sin\Theta}{4\pi \omega r_\mathrm{A} r_\mathrm{B}} \sum^\infty_{n=1} (2n+1)B^N_n(\omega) P'_n(\gamma) Q_n^{(3)},
        \\
        G_{\theta \theta}^{(1)}(\mathbf{r}_\mathrm{A},\mathbf{r}_\mathrm{B},\omega) &= -\frac{i\omega}{4\pi } \sum^\infty_{n=1} \frac{2n+1}{n(n+1)} \left[ B^M_n(\omega) P'_n(\gamma) Q_n^{(1)} + \frac{c^2 B^N_n(\omega)}{\omega^2 r_\mathrm{A} r_\mathrm{B}} F_n(\gamma) Q_n^{(4)} \right],
        \\
        G_{\phi \phi}^{(1)}(\mathbf{r}_\mathrm{A},\mathbf{r}_\mathrm{B},\omega) &= -\frac{i\omega}{4\pi} \sum^\infty_{n=1} \frac{2n+1}{n(n+1)} \left[ B^M_n(\omega) F_n(\gamma) Q_n^{(1)} + \frac{c^2 B^N_n(\omega)}{\omega^2 r_\mathrm{A} r_\mathrm{B}} P'_n(\gamma) Q_n^{(4)} \right],
    \end{align}
\end{subequations}
where we have defined,
\begin{subequations}
    \begin{align}
         Q^{(1)}_n =& h_n^{(1)}(r_\mathrm{A}\omega) h_n^{(1)}(r_\mathrm{B}\omega), \\
          Q_n^{(2)} =& h_n^{(1)}(r_\mathrm{A}\omega) \left[z h_n^{(1)}(z) \right]'_{z=r_\mathrm{B}\omega},\\
        Q_n^{(3)} =& h_n^{(1)}(r_\mathrm{B}\omega) \left[y h_n^{(1)}(y) \right]'_{y=r_\mathrm{A}\omega},\\
        Q_n^{(4)} =& \left[y h_n^{(1)}(y) \right]'_{y=r_\mathrm{A}\omega} \left[z h_n^{(1)}(z) \right]'_{z=r_\mathrm{B}\omega}\\
        F_n(x) =& n(n+1)P_n(x)-xP'_n(x).
    \end{align}
\end{subequations}
This is in agreement with results in \cite{buhmann2012Book1,safari_interatomic_2008}.

We also need to express the vacuum Green's tensor in our chosen spherical coordinate system. To do this we begin by writing the all-distances Cartesian expression \eqref{fullVacuumG} as;
\begin{equation}
    \tens{G}^{(0)}(\mathbf{r},\mathbf{r}',\omega)=- \frac{ e^{i\omega\rho}}{4\pi \omega^2 \rho^3}  \left[ a\left( -i\rho \omega \right) \mathbb{I} 
     - b\left( -i\rho \omega \right) \mathbf{e}_{\rho} \otimes \mathbf{e}_{\rho} \right], \label{eq:G0full}
\end{equation}
defined $a(x) \equiv 1+x+x^2$ and $b(x) \equiv 3+3x+x^2$. Applying to our system of two bodies at $\mathbf{r}_\mathrm{A}$ and $\mathbf{r}_\mathrm{B}$ in the $xz$ plane, where each are either side of the $z$ axis, their separation is found to be $\rho_\mathrm{AB}=\sqrt{|\mathbf{r}_\mathrm{A}|^2 + |\mathbf{r}_\mathrm{B}|^2 - 2|\mathbf{r}_\mathrm{A}| |\mathbf{r}_\mathrm{B}|\cos\Theta}$, and the spherical unit vectors are,
\begin{subequations}
    \begin{align}
        \mathbf{e}_{r_\mathrm{A}} &= (\sin\theta_\mathrm{A}, 0, \cos\theta_\mathrm{A}), &  
        \mathbf{e}_{\theta_\mathrm{A}} &= (\cos\theta_\mathrm{A}, 0, -\sin\theta_\mathrm{A}), & 
        \mathbf{e}_{\phi_\mathrm{A}} &= (0, 1, 0), \\
        \mathbf{e}_{r_\mathrm{B}} &= (-\sin\theta_\mathrm{B}, 0, \cos\theta_\mathrm{B}), &  
        \mathbf{e}_{\theta_\mathrm{B}} &= (-\cos\theta_\mathrm{B}, 0, -\sin\theta_\mathrm{B}),&  
        \mathbf{e}_{\phi_\mathrm{B}} &= (0, -1,0).
    \end{align}
\end{subequations} 
Projecting the vacuum Green's tensor \eqref{eq:G0full} onto these bases, we find the non-vanishing components to be,
\begin{subequations}
    \begin{align} \label{eq:G0_Sph}
        &\tens{G}_{rr}^{(0)}(\mathbf{r}_\mathrm{A},\mathbf{r}_\mathrm{B},\omega) = - \frac{ e^{-\xi}}{4\pi \omega^2 \rho_\mathrm{AB}^3}  \left[ a\left( \xi \right) \cos\Theta
         - b\left(\xi\right) \frac{\left(r_\mathrm{A} - r_\mathrm{B}\cos\Theta \right) \left(r_\mathrm{A}\cos\Theta - r_\mathrm{B} \right)}{\rho_\mathrm{AB}^2} \right],
         \\
        &\tens{G}_{r\theta}^{(0)}(\mathbf{r}_\mathrm{A},\mathbf{r}_\mathrm{B},\omega) =  \frac{ e^{-\xi}}{4\pi \omega^2 \rho_\mathrm{AB}^3}  \left[ a\left(\xi \right) \sin\Theta
        - b\left(\xi\right) \frac{\left(r_\mathrm{A} - r_\mathrm{B}\cos\Theta \right) r_\mathrm{A}\sin\Theta }{\rho_\mathrm{AB}^2} \right],
        \\
        &\tens{G}_{\theta r}^{(0)}(\mathbf{r}_\mathrm{A},\mathbf{r}_\mathrm{B},\omega) =  \frac{ e^{-\xi}}{4\pi \omega^2 \rho_\mathrm{AB}^3}  \left[ a\left(\xi\right) \sin\Theta
         + b\left(\xi\right) \frac{\left(r_\mathrm{A}\cos\Theta - r_\mathrm{B} \right) r_\mathrm{B}\sin\Theta }{\rho_\mathrm{AB}^2} \right],
         \\
         &\tens{G}_{\theta\theta}^{(0)}(\mathbf{r}_\mathrm{A},\mathbf{r}_\mathrm{B},\omega) = \frac{ e^{-\xi}}{4\pi \omega^2 \rho_\mathrm{AB}^3}  \left[ a\left(\xi\right) \cos\Theta
         - b\left(\xi\right) \frac{r_\mathrm{A} r_\mathrm{B}\sin^2\Theta }{\rho_\mathrm{AB}^2} \right],
         \\
         &\tens{G}_{\phi\phi}^{(0)}(\mathbf{r}_\mathrm{A},\mathbf{r}_\mathrm{B},\omega)= \frac{ e^{-\xi}}{4\pi \omega^2 \rho_\mathrm{AB}^3} a\left(\xi \right),
    \end{align}
\end{subequations}
where we have defined $\xi \equiv -i\omega\rho_\mathrm{AB}$. This result is in agreement with \cite{buhmann2012Book1,safari_interatomic_2008}.

\section{Material losses}

\begin{figure}
    \centering
    \includegraphics[width=0.5\textwidth]{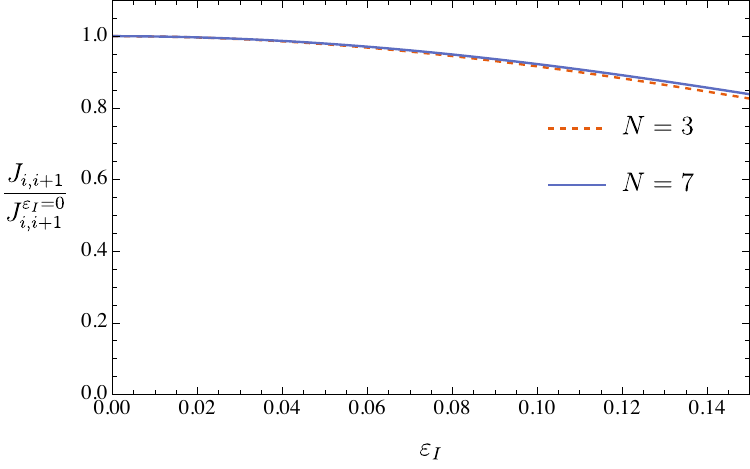}
    \caption{Change in the nearest neighbour coupling constant $J_{i,i+1}$ due to the introduction of absorption within the sphere via an imaginary part of the permittivity $\varepsilon_I$. Values are shown for $N=3$ and $N=7$ emitters in the ring -- all other choices of $3<N<7$ fall between the two lines shown. }
    \label{fig:lossPlot}
\end{figure}

In the main text we assume a real permittivity of $\varepsilon = -2.37$. However, all materials exhibit some degree of loss, modelled via the permittivity having an imaginary part. To justify our assumption of real permittivity we have investigated the effect of adding a small  (as is relevant for plasmonic materials) imaginary part to the permittivity upon nearest-neighbor values of $J_{ij}$ in the vicinity of the sphere. As shown in Fig \ref{fig:lossPlot}, for imaginary permittivity at the level of a few percent of the magnitude of the real part of the permittivity, the coupling matrix element retains approximately the same magnitude as if absorption were ignored, and --- more importantly --- comfortably retains its sign. This conclusion is also robust to changing the number of emitters in the ring, justifying the assumption of real permittivities for all systems considered in the main text. 
\\~\\

\section{Bloch-Redfield equations}
The usual second-order Markovian master equation takes the form~\cite{TheoryOQSBook} 
\begin{gather}
    \frac{d}{dt}\rho_S(t) = -i [H_S,\rho_S(t)] \\- \int^\infty_0 d\tau \text{Tr}_B\left\{\left[ H_I(t),[H_I(t-\tau),\rho_S(t)\rho_B]\right]\right\},\nonumber
\end{gather}
with $H_I(t) = H_{I,\text{opt}}(t)+H_{I,\text{vib}}(t)$, where we have used that  $S(t)$ denotes an operator $S$ in the interaction picture. 
Noting that we can decompose our interaction Hamiltonian $H_I(t) = \sum_\alpha A_\alpha(t)\otimes B_\alpha(t)$, where operators $A_\alpha$ and $B_\alpha$ act on the ring system and baths, respectively, we can rewrite the non-unitary second term on the right-hand side of the Bloch-Redfield master equation as 
\begin{gather}
    \mathcal{D} \rho_S(t)= \sum_{n,m,\alpha,\beta} \Gamma_{\alpha\beta}(\omega_{nm})[A^\alpha_m(\omega_m)\rho_S(t) A_n^{\beta\dag}(\omega_n) - A^{\beta\dag}_n(\omega_n)A^\alpha_m(\omega_m)\rho_S(t) +\text{H.c.}],
\end{gather}
where H.c.~denotes the Hermitian conjugate, $A^\alpha_n$ are projections of the system operator $A_\alpha$ onto the $n$th eigenstate of the system Hamiltonian $H_S$, and $\Gamma_{\alpha\beta}$ are the rates associated with the transition between eigenstates with energy $\omega_n$ and $\omega_m$ due to the interactions $B_\alpha$ and $B_\beta$. The rate is then determined by the difference between the two energy levels $\omega_{mn} = \omega_m-\omega_n$. These rate functions are related to the bath operators by Fourier transforms of the two-time correlations
\begin{equation}
    \Gamma_{\alpha\beta}(\omega) = \int^\infty_0  e^{i\omega s}\langle B_\beta^{\dag}(t) B_\alpha(t-s) \rangle ds.
\end{equation}
Under the assumption of thermalised photon and phonon environments, these correlation functions decompose into linear combinations of rates for both the optical and vibrational baths and   are of the form 
\begin{equation}
    \Gamma^\mu_{\alpha\beta}(\omega) =  \frac{1}{2}\gamma_{\alpha\beta}^\mu(\omega) +  i S^\mu_{\alpha\beta}(\omega),
\end{equation}
where $\mu \in \{\text{vib , opt}\}$. The real-valued component $\gamma^\mu_{nm}$ is associated with the rate of the transition and the imaginary component $S^\mu_{nm}$ is associated with a Lamb shift which is typically neglected due to being negligibly small.
The remaining rate contribution then takes the form 
\begin{equation}
\gamma^\mu_{nm}(\omega) = J^\mu(\omega)N(\omega),
\end{equation}
where $J^\mu(\omega)$ is the spectral density associated with the bath and $N(\omega)$ is related to the Bose-Einstein distribution $n(\omega)$ for the bosonic vibrational and optical bath modes, as 
\begin{equation}
    N(\omega) = \begin{cases}
         (1+n(\omega)),\hspace{5pt} \omega \geq 0,\\
         n(\omega), \hspace{30pt}\omega < 0.
         
    \end{cases}
\end{equation}
These relations ensure detailed balance for our transitions. For simplicity and to study and expose the superabsorption mechanism unperturbed by environmental specifics, we take for both the optical and vibrational spectral densities a flat spectrum 
\begin{equation}
    J^\mu(\omega) =  \begin{cases}
        \gamma_\text{vib} = 10~\text{meV},\\
        \gamma_\text{opt} = 1~\mu\text{eV}.
    \end{cases}
\end{equation}
Due to the decomposition of the correlations into optical and vibrational contributions, in keeping with our second-order treatment we can decompose the Redfield dissipator into an additive linear combination of optical and vibrational dissipators as 
\begin{equation}
    \mathcal{D} \rho_S  = (\mathcal{D}_\text{vib} + \mathcal{D}_{\text{opt}}) \rho_S.
\end{equation}

\section{Trap Quantum Heat Engine Dissipators}
The abstract load of the trap is coupled to the ring-like system via incoherent processes for extraction. This is implemented through an additional dissipator of Lindblad form in the master equation 
\begin{equation}
    \mathcal{D}_X\rho_S = \gamma_X( X\rho_S X^\dag - \frac{1}{2}\{X^\dag X, \rho_S\}),
\end{equation}
    
where $\gamma_X$ is the rate of extraction from the ring to the trap and the Lindblad operator $X$ defining this incoherent transport is 
\begin{equation}
    X = \sum_{i=1}^N\sigma_-^{(i)}\otimes \sigma_+^{(t)}.
\end{equation}

Additionally, we introduce a trap decay dissipator, causing incoherent transitions from the excited state of the trap to its ground state of the form  
\begin{equation}
    \mathcal{D}_t\rho_S = \Gamma_t (\sigma_-^{(t)}\rho_S \sigma_+^{(t)} - \frac{1}{2}\{\sigma_+^{(t)}\sigma_-^{(t)},\rho_S\}).
\end{equation}

The reduced density matrix therefore obeys the master equation
\begin{equation}
    \frac{d}{dt}\rho_S =-i[H_S,\rho_S] +(\mathcal{D}_\text{opt}+\mathcal{D}_\text{vib}+\mathcal{D}_X + \mathcal{D}_t)\rho_S.
    \label{eqn:FullMasterEquation}
\end{equation}

\section{Power output from quantum heat engine}
From the abstract load we are able to calculate a power output. We calculate the current of the quantum heat engine given by 
\begin{equation}
    I =  e\Gamma_t\langle \rho_{t,e}\rangle,
\end{equation}
where $\langle \rho_{t,e}\rangle$ is the expectation value of the steady state population for the excited state of the trap. Using this formalism we also calculate the potential difference generated as given by 
\begin{equation}
    eV = \hbar\omega_t + k_B T_{\text{vib}} \ln{\left[\frac{\langle \rho_{t,e}\rangle}{\langle \rho_{t,g}\rangle}\right]},
\end{equation}
where $T_\text{vib}$ is the temperature of the vibrational baths and $\langle \rho_{t,g}\rangle$ is the expectation of the steady state population for the trap's ground state~\cite{Scully2011}. From this, we compute the power output of the solar absorber at the steady state by $P = IV$. Steady states are calculated by finding the null-space of Eq.~\eqref{eqn:FullMasterEquation}.
By allowing for $\Gamma_t$ to vary, we may find the optimal power output for each configuration of the ring system, such that $P_\text{max}=\text{max}_{\Gamma_t}\{I(\Gamma_t)V(\Gamma_t)\}$. 

\section{Validity of Bloch-Redfield Equations}
As previously mentioned, the Bloch-Redfield equation utilises weak coupling and Markovian approximations in its derivation. The Bloch-Redfield equations can be derived from the exact Nakajima-Zwanzig equations, by taking terms up to second order and extending the time integral to infinity. This second step is justified as we are only interested in steady-state power outputs and all non-Markovian elements are lost. Secondly, we can justify the 2nd order approach by comparing with the variational polaron transformation approach shown in \cite{Pollock_2013}, capable of modelling weak, intermediate, and strongly coupled systems. To this end, we use a more typical form for the spectral density of the vibrational bath, 
\begin{equation}
    J^\text{vib}(\omega) = \alpha \frac{\omega^3}{\omega_C^2}\exp\left\{-\left(\frac{\omega}{\omega_C}\right)^2\right\}.
\end{equation}
This super-Ohmic spectral density with $\omega_C=100$meV and $\alpha=1.2$ leads to a spectral density of comparable magnitude to the original flat band for all relevant frequencies. From this, we can follow the procedure in \cite{Pollock_2013} to determine how much displacement we anticipate for the phonon modes by interacting with the system. Doing so yields a displacement distribution of the form 
\begin{equation}
    F(\omega)  = \frac{\omega}{\omega+\kappa \coth{(\beta\omega/2})},
\end{equation}
where $\kappa$ is the variational parameter we are solving for. For the 6-site system the value of $\kappa=422$~meV within the single excitation manifold. As this is significantly larger than the energies associated with our spectral density, it follows that the weak coupling framing taken in the results shown is sufficient. From these two points, namely, interest in the steady state and a low value for the displacement distribution, the Bloch-Redfield master equation is well justified for our uses. 

\section{Parameters used for simulations}
The parameters used for the simulations presented in this paper have been chosen to be representative of typical molecular chromophores and are outlined in Table.~\ref{tab:params}.
\begin{table}[h!]
    \centering
\begin{tabular}{ |c| c|}
\hline
 Parameter & Value  \\ 
 \hline
 $T_\text{vib}$ & $300$K  \\  
 $T_\text{opt}$ & $5800$K   \\  
 $\gamma_\text{opt}$ & $1\mu$eV\\
 $\gamma_\text{vib}$ & $10^3\gamma_\text{opt}$\\
 $\gamma_X$ & $\gamma_\text{opt}$\\
 $\omega_0$ & $1.8$eV \\
 $r_{\text{NN}}$ & $2.5$nm \\
 $r_{\text{S}}$ & $(r_R - 1)$nm \\
 $\epsilon$ & $-2.37$\\
  \hline
\end{tabular}  
    \caption{Parameters used in the simulations for the guide-slide systems.}
    \label{tab:params}
\end{table}

\section{Reinitialisation and Photonic band-gap}
\label{sec:Reinit}

\begin{figure}
    \centering
    \includegraphics[width=0.6\textwidth]{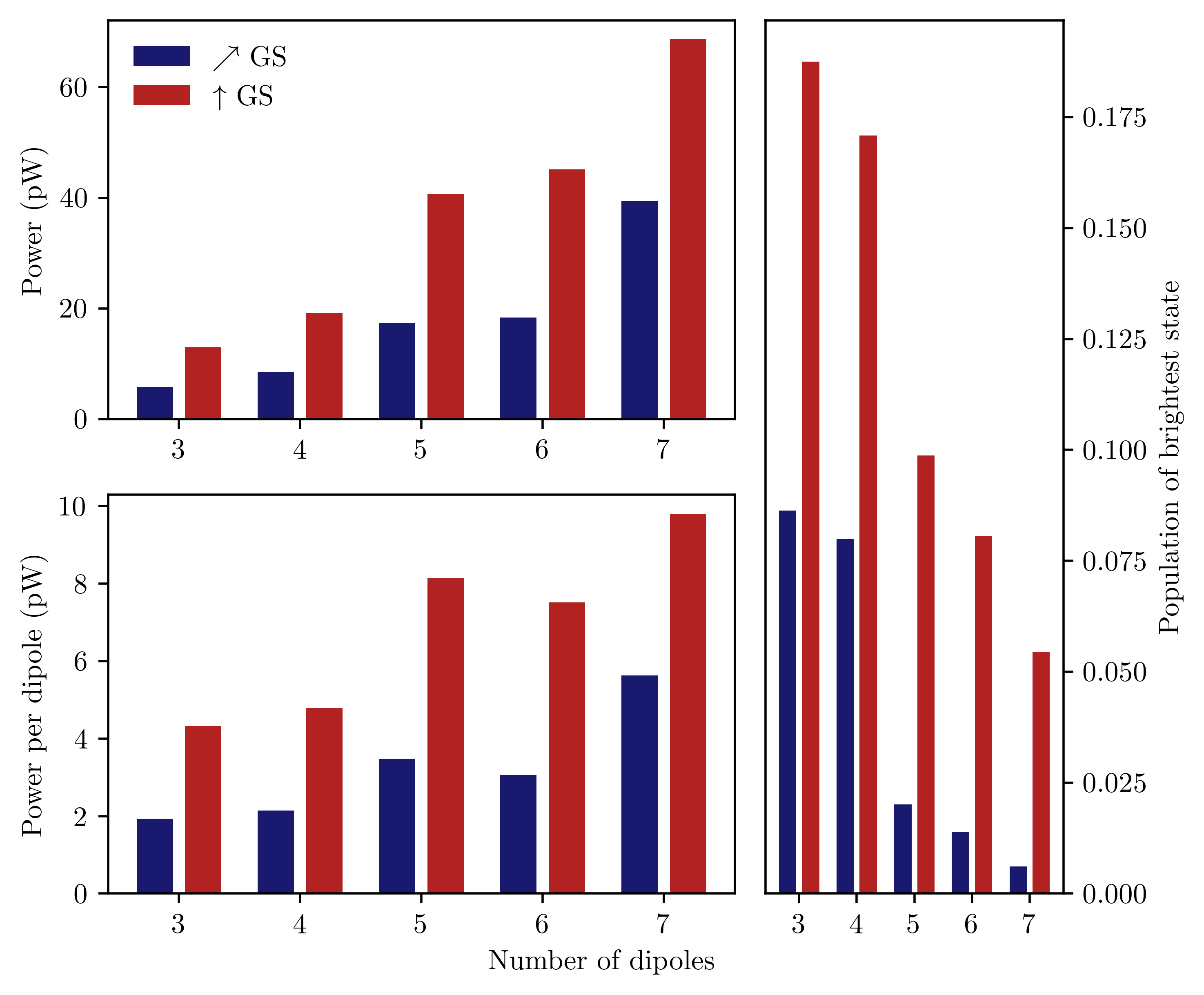}
    \caption{Scaling of the solar cell power output for the configuration of the plasmon-enabled guide slide, as well as the previously developed tilted guide slide, with driving and confinement within a photonic band gap. We also show the population of the brightest state for each $N$. As previously observed \cite{Brown2019}, odd ring sizes are favoured, featuring higher accessible oscillator strength lower on the Dicke ladder. This effect is more pronounced under active reinitialisation compared to the results in the main text.}
    \label{fig:NScalingReinit}
\end{figure}

We now compare the results for utilising the reinitialisation scheme and photonic band gap suppression outlined in \cite{Brown2019}. This reinitialisation scheme introduces a new Lindblad term 
\begin{equation}
   \mathcal{D}_R\rho_S = \gamma_R( R\rho_S R^\dag - \frac{1}{2}\{R^\dag R, \rho_S\}),
\end{equation}
where $R$ takes all Dicke ladder states below the target state (immediately below half inversion on the Dicke ladder) to this target state. We utilise a pumping rate $\gamma_R = 0.1\gamma_o$, but have observed qualitatively similar behaviour for different choices of this rate. The photonic bandgap is implemented by rescaling the optical transitions such that 
\begin{equation}
       J^{\text{opt}}(\omega) = \begin{cases}
         \gamma_\text{opt},\hspace{23pt} |\omega| \geq \omega_T,\\
         0.01\gamma_\text{opt}, \hspace{5pt}|\omega| < \omega_T.
         
    \end{cases}
\end{equation}
$\omega_T$ is the target transition from immediately below the half-inversion manifold to the next lowest state on the Dicke ladder. These two effects aim to maintain the superabsorber near the half-inversion regime where the transition dipoles are maximally enhanced for absorption.  
In Fig.~\ref{fig:NScalingReinit}, we have the results of these simulations for varying $N$. Both systems benefit substantially from the reinitialisation, however, the plasmon-enabled guide-slide still outperforms the tilted guide slide. This is well understood when we consider the localisation on the brightest state in the system at the half-inversion point, which is consistently substantially higher for the plasmon-enabled guide slide system. Note that the population of this site does decrease with $N$. This is due to the number of states in the system increasing exponentially. However, more states on the Dicke ladder `nearby' to the half-inversion state will have near $N^2$ scaling of their transition dipole strengths. 

\section{Purcell Enhancement of Optical Decay Rates}
\label{sec:Purcell}
Fundamentally, structuring the electromagnetic field modes around the transition frequency affects dipolar couplings and leads to a modulation of the optical decay rates due to the Purcell effect. In the main text, we have neglected such an effect to ensure that the effects we are seeing are solely due to super-absorbing enhancements. We now consider these effects and show that they do not change the narrative of the manuscript, and in fact may be beneficial for super-linear scaling of the output power, although in this case partially attributable to individual in addition to collective enhancement.

The modification of the optical decay rates due to the structuring of the electromagnetic modes around the sphere has been calculated as ~\cite{Leung1988}
\begin{equation}
    \gamma_\text{opt} = \gamma_\text{opt,free}\left[1-\left(\frac{R_s}{R_r}\right)^3 \frac{\epsilon-1}{\epsilon+2}\right]^2.
\end{equation}
For the current setup we have, with $\epsilon = -2.37$, we can note that as the size of the ring increases the ratio between the sphere and ring radius tends towards unity, and thus 
\begin{equation}
    \gamma_\text{opt}^\text{sat} = \lim_{R_s\rightarrow R_r}\gamma_\text{opt} = \gamma_\text{opt,free}\left[1-\frac{\epsilon-1}{\epsilon+2}\right]^2 \approx 33.7 \gamma_\text{opt,free}.
\end{equation}
And in fact $\gamma_\text{opt}>1$ for all $N>6$. Thus, we should anticipate a greater super-linear scaling when accounting for the $N$-dependent Purcell enhancement. Furthermore, we have also checked the variation of the local density of states around the transition frequencies and have shown it to be flat. As such, by introducing the optical structuring of the environment we have not added non-Markovian effects that may invalidate the approach taken. 

\begin{figure}
    \centering
    \includegraphics[width=0.45\textwidth]{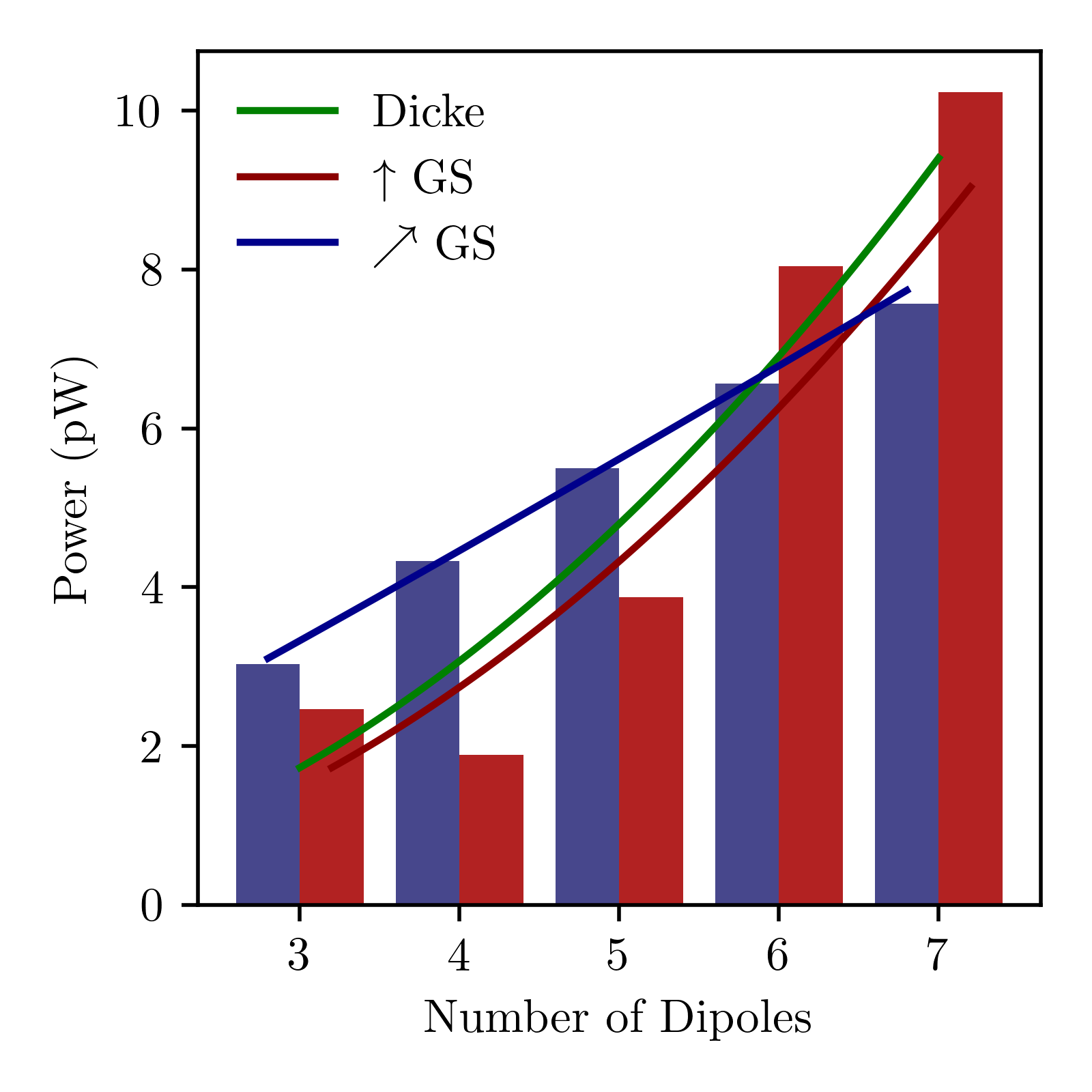}
    \caption{Scaling of the solar cell power output for the configuration of the plasmon-enabled guide slide, as well as the previously developed tilted guide slide.  The super linear scaling for each configuration $P = \alpha N^m$. The parallel sphere guide slide has $m=1.95$ and the tilted guide slide $m=1.08$. The theoretical maximum Dicke scaling of $N^2$ is also plotted. }
    \label{fig:NScalingPurcell}
\end{figure}

Fig.~\ref{fig:NScalingPurcell} shows simulation results analogue to the main text but with the Purcell enhancement effect included. As predicted, we observe a further enhancement of the super-absorption, now with an enhanced exponent of $m=1.95$. However, we note this is due to the combination of different effects, and thus only coincidentally close to the theoretical superabsorption Dicke limit ($m=2$). 

\section{Effects of dephasing on superabsorption}
\begin{figure}
    \centering
    \begin{subfigure}{.45\textwidth}
        \includegraphics[width=\linewidth]{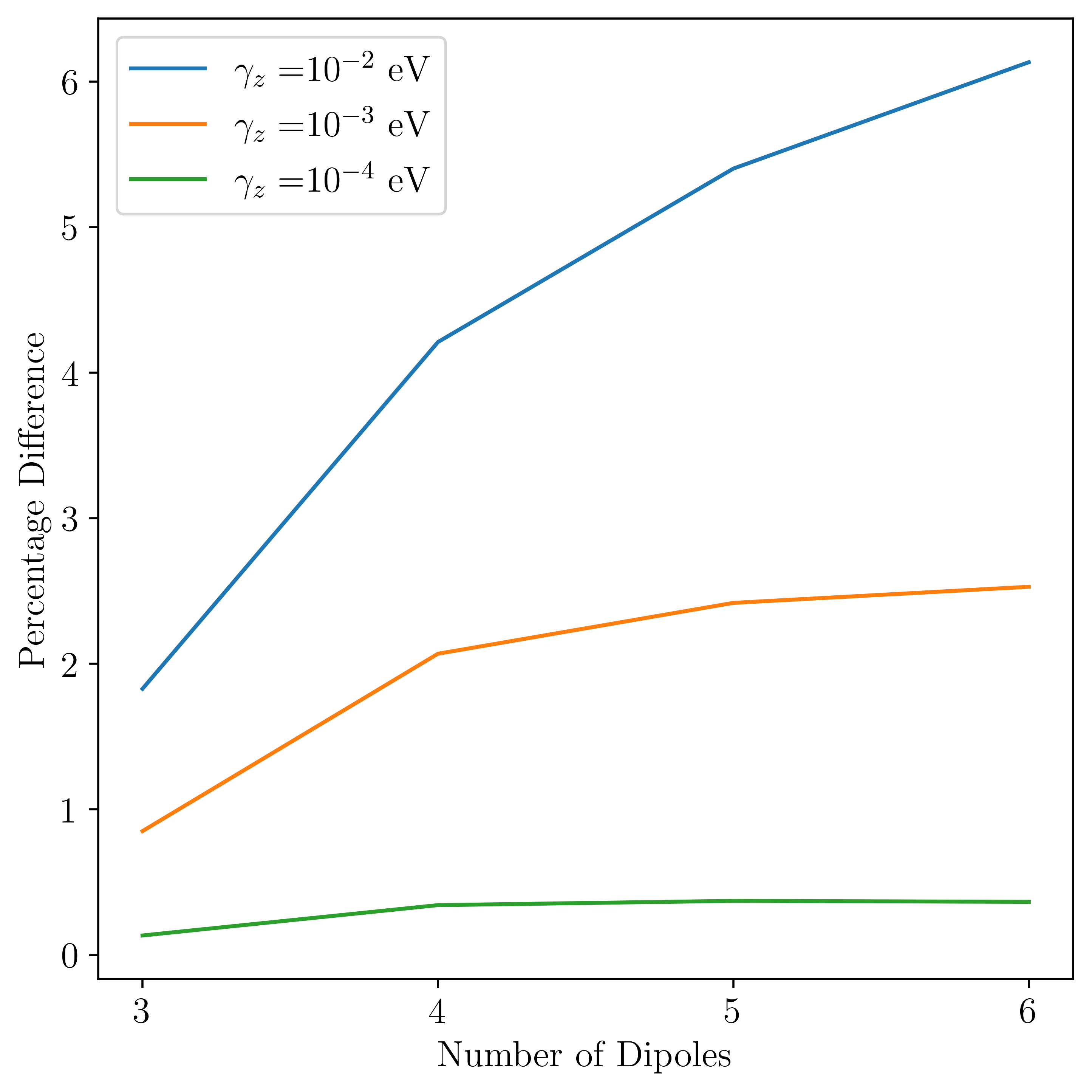}
        \caption{Surface plasmon-enabled guide-slide}
    \end{subfigure}
        \begin{subfigure}{.45\textwidth}
        \includegraphics[width=\linewidth]{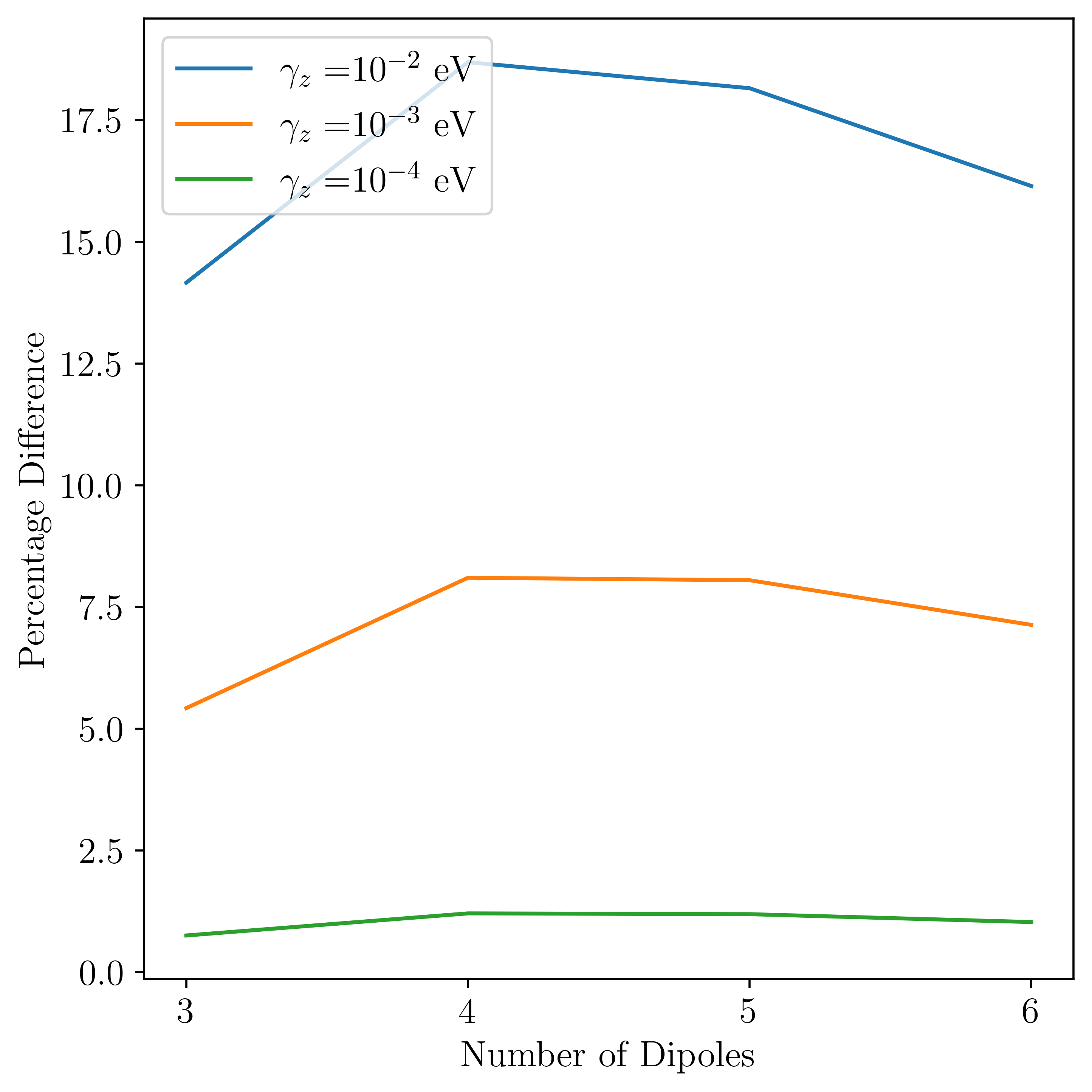}
        \caption{Tilted guide-slide}
    \end{subfigure}
    \caption{Percentage reduction in power outputs of the superabsorbers with varying rates of dephasing $\gamma_z$ compared with the zero dephasing case presented in the main text for different number of dipoles in the ring. }
    \label{fig:Deph}
\end{figure}

In many condensed matter systems environmental noise can lead to dephasing of energy levels of atomic systems. Whilst our model already accounts for vibrational effects which often represent the predominant source of dephasing, other contributions may be present, or one might see additional vibrational dephasing beyond what is predicting by our second order perturbative phonon treatment. For this reason, we here explore how deleterious dephasing impacts the ability for the discussed systems to perform superabsorption. This is achieved by adding in additional Lindblad dissipation terms into the master equation. We utilise standard dephasing $\sigma_z^{(i)}$ Lindblad operators with a rate $\gamma_z$ such that the additional dissipator takes the form

\begin{equation}
 \mathcal{D}_{\text{deph}} \rho_S = \sum_{i=1}^N \mathcal{D}_{i,\text{deph}} \rho_S,
\end{equation}
with 
\begin{equation}
    \mathcal{D}_{i,\text{deph}} \rho_S =  \gamma_z( \sigma_z^{(i)}\rho_S  \sigma_z^{(i)}-  \rho_S).
\end{equation}
In Fig.~\ref{fig:Deph}, we show the percentage difference between the simulations for different magnitudes of dephasing rate $\gamma_z$ and the zero dephasing case outlined in the main text. We vary these magnitudes around the phonon decay rates. What becomes readily apparent is that dephasing has a marginal negative effect on the power output of the newly described superabsorber even for strong dephasing $\gamma_z = 10\gamma_\text{vib}$ . This is not entirely surprising as the design of the superabsorber was such as to leverage environmental noise to enhance the absorption. One can consider such dephasing as having a similar effect to the vibrational dissipation but with a much larger effective temperature. As such dephasing will transfer excitations between dark and bright states, however the trap extraction process breaks the detailed balance leading to a still out-of-equilibrium situation. Thus dephasing does not destroy the superabsorbing behaviour. We do note that this has a $N$-dependent effect and does slightly reduce the super-extensive scaling of the power outputs. We also note that in the new plasmon-enabled guide-slide system, the effects are far less severe for each level of dephasing compared with the tilted guide-slide, further exhibiting its benefits.

\section{Effects of detuning on superabsorption}
\begin{figure}
    \centering
    \begin{subfigure}{.45\textwidth}
        \includegraphics[width=\linewidth]{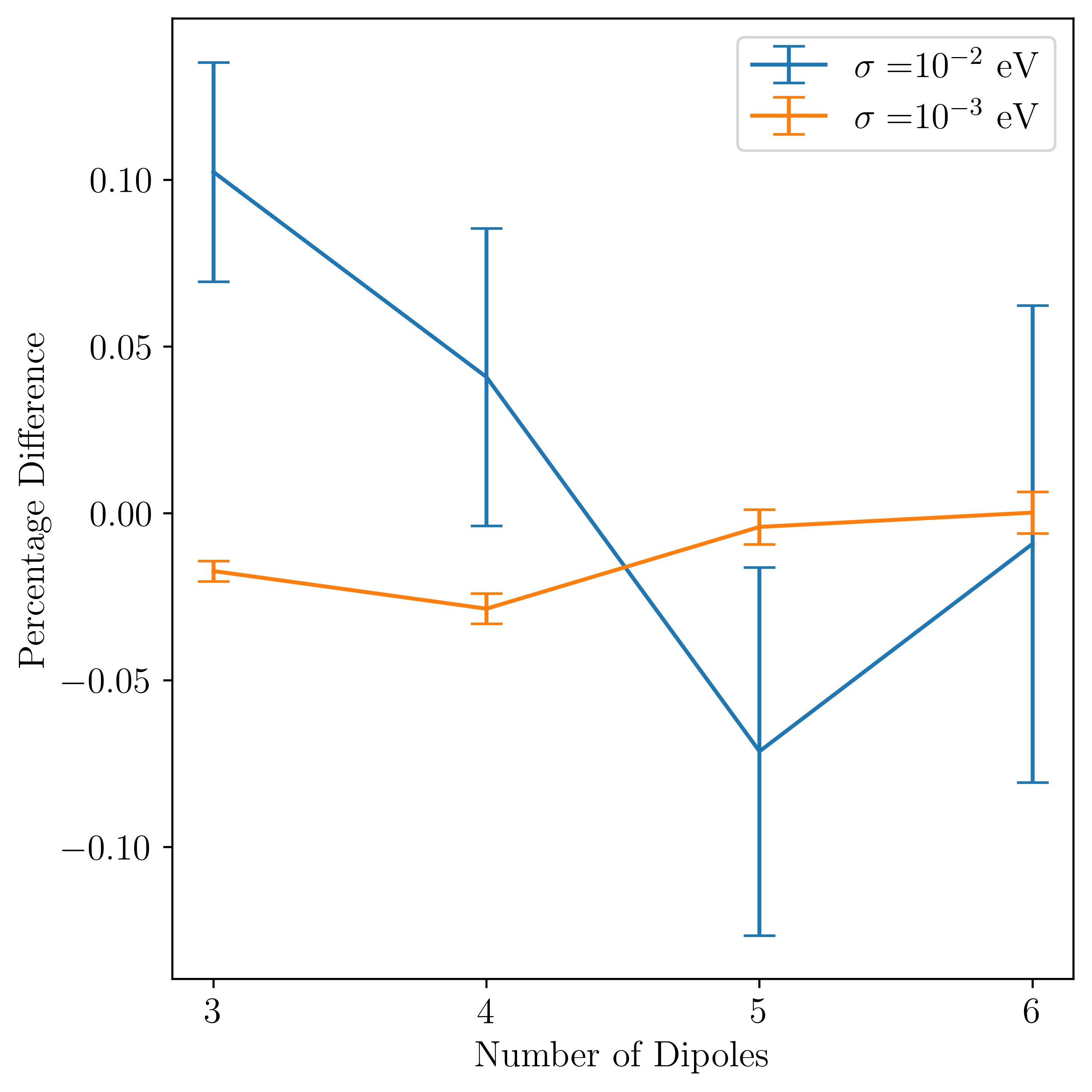}
        \caption{Surface plasmon-enabled guide-slide}
    \end{subfigure}
        \begin{subfigure}{.45\textwidth}
        \includegraphics[width=\linewidth]{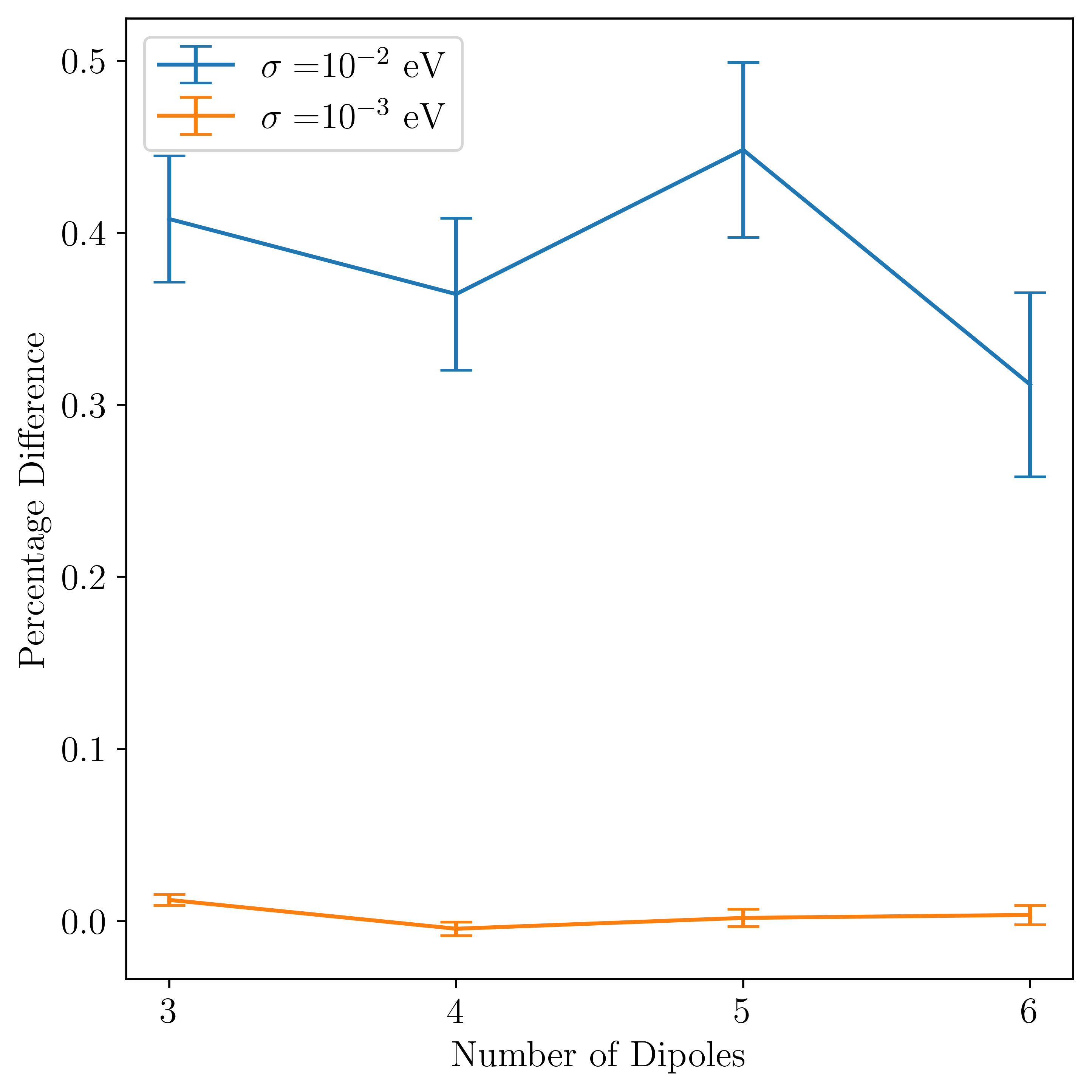}
        \caption{Tilted guide-slide}
    \end{subfigure}
    \caption{Percentage difference  in power outputs of the superabsorbers with varying rates of detuning of energy levels. Positives numbers denote a reduction whilst negative ones indicate an increase. Results show the mean of 100 iterations for each $\sigma$  alongside error bars representing the standard deviation of the sample. All results are  relative to the zero detuning results presented in the main text for different number of dipoles in the ring. 
    }
    \label{fig:Detuning}
\end{figure}

Similarly to dephasing, a common feature of realistic systems are the effects of static disorder, leading to shifts in the energy levels of the constituent dipoles. Here we look at how such random detunings impact the ability for these systems to superabsorb. In Fig.~\ref{fig:Detuning}, we show the results of sampling 100 different energetically disordered systems by shifting the transition energies of each of the sites by a normally distributed value $\epsilon_i$, with varying standard deviation $\sigma$. What we see is that the plasmon enabled guideslide is very robust to moderate amounts of static noise ($<1\%$ losses), and even for some values of noise may exhibit small enhancements for the smaller ring systems. A similar effect has been observed previously for in Ref.~\cite{Higgins:2017aa}, and may be due to new energetic pathways opening up in the disordered system allowing for a greater power output. We also note that the parallel guide-slide is again more robust to such fluctuations compared with the tilted guide-slide. This is, in part, due to the strong dipole-dipole coupling ($\approx 40$ meV) generated by the plasmonic coupling that makes the delocalised bright states robust to fluctuations. For much larger fluctuations other physical processes would need to be considered, such as corrections to adequately capture the frequency dependence of both optical and vibrational spectral densities across such larger frequency ranges, but this is outside the scope of the current study.  

